# Perfect Lattice Actions for Quarks and Gluons [*]

W. Bietenholz and U.-J. Wiese

Center for Theoretical Physics,
Laboratory for Nuclear Science, and Department of Physics
Massachusetts Institute of Technology (MIT)
Cambridge, Massachusetts 02139, U.S.A.

MIT Preprint, CTP 2475

October 23, 1995

## Abstract

We use perturbation theory to construct perfect lattice actions for quarks and gluons. The renormalized trajectory for free massive quarks is identified by blocking directly from the continuum. We tune a parameter in the renormalization group transformation such that for 1-d configurations the perfect action reduces to the nearest neighbor Wilson fermion action. The fixed point action for free gluons is also obtained by blocking from the continuum. For 2-d configurations it reduces to the standard plaquette action. Classically perfect quark and gluon fields, quark-gluon composite operators and vector and axial vector currents are constructed as well. Also the quark-antiquark potential is derived from the classically perfect Polyakov loop. The quark-gluon and 3-gluon perfect vertex functions are determined to leading order in the gauge coupling. We also construct a new block factor $n$ renormalization group transformation for QCD that allows to extend our results beyond perturbation theory. For weak fields it leads to the same perfect action as blocking from the continuum. For arbitrarily strong 2-d Abelian gauge fields the Manton plaquette action is classically perfect for this transformation.

---

[*]This work is supported in part by funds provided by the U.S. Department of Energy (D.O.E.) under cooperative research agreement DE-FC02-94ER40818.

# 1 Introduction

Cut-off effects are the major source of systematic errors in numerical simulations of lattice QCD. For the standard Wilson action these effects are of the order of the lattice spacing. Hence they vanish rather slowly as the continuum limit is approached. Symanzik's perturbative improvement program systematically eliminates cut-off effects by introducing irrelevant higher-dimensional operators in the lattice action [1]. For QCD this program was realized by Sheikholeslami and Wohlert [2]. Their so-called 'clover action' is now often used in numerical simulations. A related attempt to damp cut-off effects uses the concept of tadpole improvement [3].

Recently Hasenfratz and Niedermayer proposed a new approach to eliminate cut-off effects by using nonperturbative approximations of 'perfect actions' [4]. By definition a perfect action is free of any cut-off artifacts. It is located on a renormalized trajectory emanating from a fixed point on the critical surface. The crucial observation of [4] is that the construction of the fixed point action in an asymptotically free theory is a classical field theory problem that can be solved nonperturbatively. In fact, the fixed point action is a classically perfect action. Using this action even in the quantum theory Hasenfratz and Niedermayer observed that cut-off effects are practically eliminated in the 2-d $O(3)$ nonlinear $\sigma$-model at correlation lengths as small as 3 lattice spacings. The action has been combined with the classically perfect topological charge in a study of the scaling properties of the topological susceptibility [5]. Recently the renormalized trajectory for a $\delta$-function renormalization group transformation was investigated for the $O(N)$ model in [6]. Wieczerkowski and Xylander combined the concepts of perfect and Symanzik improved lattice actions in a study of the 2-d hierarchical $O(N)$ nonlinear $\sigma$-model [7]. For the Gross-Neveu model in the large $N$ limit it has been shown analytically that the fixed point action is exactly perfect at the quantum level even at arbitrarily small correlation length [8]. Also it has been realized that a classically perfect action is automatically 1-loop quantum perfect [9]. Finally the perfect action approach has been applied successfully to pure $SU(3)$ gauge theory [10]. Other constructions of perfect lattice actions for gauge theories are contained in [11]. A more formal discussion of block transformations for gauge fields is given in [12], and early numerical work on $SU(2)$ can be found in [13].

In the present paper we concentrate on the inclusion of the quarks. Using perturbation theory we construct a lattice QCD action that is perfect to first order in the gauge coupling for arbitrary fermion mass. This is an essential step towards a fully nonperturbative perfect action for QCD. Since nonperturbative effects are not yet included our action is not intended to be used in simulations of light quarks. For heavy quarks, however, the leading cut-off effects are related to the heavy fermion mass. In our approach these effects are completely eliminated to first order in the gauge coupling. Since the gauge coupling is weak at the scale of a heavy fermion mass, our perturbatively perfect action is promising for numerical simulations of heavy quarks. The inclusion of nonperturbative effects will be discussed in a separate publication [14]. Then we will be able to treat light quarks as well. The question of locality of the action plays an important role throughout this paper. We distinguish three types of locality. An action is called 'ultralocal' if its couplings are limited to a finite set of neighbors. It is called 'local' if the couplings decay exponentially at large distances, and 'nonlocal' if they decay only powerlike. The main technical tool of this paper is a renormalization group transformation that defines lattice fields by blocking directly from the continuum. In section 2 we discuss this transformation for free massive quarks. The perfect action is determined and a parameter in the renormalization group transformation is tuned such that the action becomes ultralocal for 1-d configurations. In section 3 we apply the technique of blocking from the continuum to free gluons. Again the parameters in the renormalization group transformation are optimized for locality of the fixed point action. For 2-d configurations the action reduces to the standard ultralocal plaquette action. In section 4 classically perfect quark and gluon fields as well as composite operators and currents are constructed. Also the static quark-antiquark potential is derived from the classically perfect Polyakov loop, and it is compared with the potentials of the standard Polyakov loop and of the continuum theory. In section 5 quarks and gluons are coupled, and perfect quark-gluon and 3-gluon vertex functions are derived to first order in the gauge coupling. Section 6 contains the construction of a nonperturbative block factor $n$ renormalization group transformation that is (upon iteration) equivalent to blocking from the continuum for weak fields. This enables the extension of our results beyond perturbation theory. The new transformation has attractive features. In particular, for arbitrarily strong 2-d Abelian gauge fields the Manton plaquette action [15] is classically perfect. Finally section 7 contains our conclusions. A synopsis of this paper has been written up in [16].

## 2  Blocking free quarks from the continuum

Block variable renormalization group transformations with a finite blocking factor are well established in lattice field theory and statistical mechanics. Iterating such transformations leads — for a suitable choice of parameters — to fixed points of the renormalization group, which are directly related to the continuum limit [17]. It has been discovered a long time ago (but has not been very well-known) that one can reach these fixed points also in one step by blocking directly from the continuum. Wilson has applied this method to scalar fields and mentioned possible generalizations [18]. A formal discussion is given in [19]. Cronjäger and Mai applied it to free fermions [20]. It has also been used in the nonlinear $\sigma$-model [4, 21]. Finally, in a recent perturbative construction of lattice chiral fermions we have derived the perfect action of fermions and gauge fields by blocking from the continuum [22].

Let us consider a quark field $\bar{\psi}, \psi$ in the continuum. We construct a lattice quark field by averaging the continuum field over hypercubes $c_x$ centered at the points $x$ of a $d$-dimensional Euclidean lattice

$$\bar{\Psi}_x = \int_{c_x} d^d y \; \bar{\psi}(y), \;\; \Psi_x = \int_{c_x} d^d y \; \psi(y). \tag{2.1}$$

As a rule, in this paper we use capital letters for lattice fields and small letters for continuum fields. The hypercubes have unit length and hence the lattice spacing is 1. All physical quantities in this paper — also those of the continuum theory — are measured in lattice units. Integrating out the continuum degrees of freedom one arrives at an effective action for the lattice fields

$$\exp(-S[\bar{\Psi},\Psi]) = \int \mathcal{D}\bar{\psi}\mathcal{D}\psi \exp(-s[\bar{\psi},\psi]) \prod_x \delta(\bar{\Psi}_x - \int_{c_x} d^d y \; \bar{\psi}(y))\delta(\Psi_x - \int_{c_x} d^d y \; \psi(y)). \tag{2.2}$$

Here $s[\bar{\psi},\psi]$ is the continuum action for free quarks. In momentum space it takes the form

$$s[\bar{\psi},\psi] = \frac{1}{(2\pi)^d} \int d^d p \; \bar{\psi}(-p)\delta^f(p)^{-1}\psi(p), \;\; \delta^f(p) = (i\gamma_\mu p_\mu + m)^{-1}, \tag{2.3}$$

where $\delta^f(p)$ is the free quark propagator in the continuum. The lattice fermion field is given by

$$\Psi(p) = \sum_{l \in \mathbf{Z}^d} \psi(p + 2\pi l)\Pi(p + 2\pi l), \;\; \Pi(p) = \prod_{\mu=1}^d \frac{\hat{p}_\mu}{p_\mu}, \;\; \hat{p}_\mu = 2\sin(p_\mu/2). \tag{2.4}$$

Note that the momentum of the lattice field is restricted to the Brillouin zone $B = ]-\pi,\pi]^d$. Due to the summation over the integer vector $l$ the lattice field is periodic over $B$. To perform the integration over the continuum fields we go to momentum space and represent the Grassmann $\delta$-function of eq.(2.2) as an integral over auxiliary lattice fermion fields $\bar{\eta},\eta$

$$\begin{aligned}\exp(-S[\bar{\Psi},\Psi]) &= \int \mathcal{D}\bar{\psi}\mathcal{D}\psi\mathcal{D}\bar{\eta}\mathcal{D}\eta \exp\left\{-\frac{1}{(2\pi)^d}\int d^d p \; \bar{\psi}(-p)\delta^f(p)^{-1}\psi(p)\right\} \\ &\times \exp\left\{\frac{1}{(2\pi)^d}\int_B d^d p \Big[[\bar{\Psi}(-p) - \sum_{l \in \mathbf{Z}^d}\bar{\psi}(-p-2\pi l)\Pi(p+2\pi l)]\eta(p)\right. \\ &\left. + \bar{\eta}(-p)[\Psi(p) - \sum_{l \in \mathbf{Z}^d}\psi(p+2\pi l)\Pi(p+2\pi l)] + a\bar{\eta}(-p)\eta(p)\Big]\right\}.\end{aligned} \tag{2.5}$$

The $\delta$-function renormalization group transformation of eq.(2.2) corresponds to $a = 0$. This transformation is chirally invariant and its fixed point (at $m = 0$) describes free lattice chiral fermions [23]. The corresponding fixed point action is nonlocal. Cronjäger has shown that its couplings decay as $r^{1-d}$ at large distances $r$ [20].

Hence it is consistent with the Nielsen-Ninomiya no-go theorem. In [22] we have shown that the nonlocality does not cause problems even in the interacting chiral gauge theory. In particular, the correct axial anomaly is reproduced on the lattice. Here we are interested in QCD which is a vector-like theory. Therefore we allow chiral breaking terms in the renormalization group transformation and hence in the fixed point action. In principle one may allow such terms also in a chiral gauge theory. [1] This possibility was explored in [16]. In fact, for numerical investigations it is essential to have a local action. Therefore in eq.(2.5) we have introduced a chiral symmetry breaking mass term $a\bar{\eta}\eta$ for the auxiliary field which smears the $\delta$-functions of eq.(2.2) to a Gaussian distribution of width $a$. For $a > 0$ the perfect action turns out to be local [23]. Later we will see explicitly that the chiral breaking introduced in the renormalization group step does not affect the spectrum of the theory.

Next we combine the integral over the Brillouin zone in eq.(2.5) with the sum over integer vectors $l$ to an integral over the entire momentum space of the continuum theory

$$\exp(-S[\bar{\Psi},\Psi]) = \int \mathcal{D}\bar{\psi}\mathcal{D}\psi\mathcal{D}\bar{\eta}\mathcal{D}\eta \exp\left\{-\frac{1}{(2\pi)^d}\int d^d p \left[\bar{\psi}(-p)\delta^f(p)^{-1}\psi(p)\right.\right.$$
$$\left.\left. + \bar{\psi}(-p)\Pi(p)\eta(p) + \bar{\eta}(-p)\psi(p)\Pi(p)\right]\right\}$$
$$\times \exp\left\{\frac{1}{(2\pi)^d}\int_B d^d p \left[\bar{\Psi}(-p)\eta(p) + \bar{\eta}(-p)\Psi(p) + a\bar{\eta}(-p)\eta(p)\right]\right\}. \quad (2.6)$$

Performing the Gaussian integral is equivalent to solving a classical equation of motion for the continuum quark field. The solution of this equation is

$$\bar{\psi}_c(-p) = -\bar{\eta}(-p)\Pi(p)\delta^f(p), \quad \psi_c(p) = -\delta^f(p)\Pi(p)\eta(p). \quad (2.7)$$

Inserting this back into the exponent yields

$$\exp(-S[\bar{\Psi},\Psi]) = \int \mathcal{D}\bar{\eta}\mathcal{D}\eta \exp\left\{\frac{1}{(2\pi)^d}\int d^d p \; \bar{\eta}(-p)\delta^f(p)\Pi(p)^2\eta(p)\right\}$$
$$\times \exp\left\{\frac{1}{(2\pi)^d}\int_B d^d p \left[\bar{\Psi}(-p)\eta(p) + \bar{\eta}(-p)\Psi(p) + a\bar{\eta}(-p)\eta(p)\right]\right\}$$
$$= \int \mathcal{D}\bar{\eta}\mathcal{D}\eta \exp\left\{\frac{1}{(2\pi)^d}\int_B d^d p \left[\bar{\Psi}(-p)\eta(p) + \bar{\eta}(-p)\Psi(p)\right.\right.$$
$$\left.\left. + \bar{\eta}(-p)[\sum_{l\in\mathbb{Z}^d}\delta^f(p+2\pi l)\Pi(p+2\pi l)^2 + a]\eta(p)\right]\right\}. \quad (2.8)$$

Performing the Gaussian integral over the auxiliary field is again equivalent to solving a classical equation of motion. In this case the solution is

$$\bar{\eta}_c(-p) = -\bar{\Psi}(-p)\Delta^f(p)^{-1}, \quad \eta_c(p) = -\Delta^f(p)^{-1}\Psi(p), \quad (2.9)$$

---

[1] We thank P. Hasenfratz for pointing this out to us.

with
$$\Delta^f(p) = \sum_{l \in \mathbf{Z}^d} \delta^f(p + 2\pi l)\Pi(p + 2\pi l)^2 + a, \tag{2.10}$$
such that together with eq.(2.7) we obtain
$$\bar{\psi}_c(-p) = \bar{\Psi}(-p)\Delta^f(p)^{-1}\Pi(p)\delta^f(p), \quad \psi_c(p) = \delta^f(p)\Pi(p)\Delta^f(p)^{-1}\Psi(p). \tag{2.11}$$
These relations between the continuum and lattice quark fields will be useful later when we construct composite operators and determine the perfect quark-gluon vertex function. Inserting eq.(2.9) back into the exponent of eq.(2.8) one obtains the perfect action
$$\begin{aligned} S[\bar{\Psi}, \Psi] &= \frac{1}{(2\pi)^d} \int_B d^d p \; \bar{\Psi}(-p)\Delta^f(p)^{-1}\Psi(p), \\ \Delta^f(p) &= \sum_{l \in \mathbf{Z}^d}[i\gamma_\mu(p_\mu + 2\pi l_\mu) + m]^{-1}\Pi(p + 2\pi l)^2 + a. \end{aligned} \tag{2.12}$$
The sum in the lattice fermion propagator $\Delta^f(p)$ converges in any dimension because $\Pi(p + 2\pi l)^2$ suppresses contributions from large $|l_\mu|$. The lattice action describes a complete renormalized trajectory parametrized by the mass $m$. At $m = 0$ the trajectory intersects the critical surface in the fixed point.

The introduction of $a > 0$ leads to a local perfect action with explicitly broken chiral symmetry even at $m = 0$. We now vary $a$ such that the action becomes as local as possible. It turns out to be sufficient to optimize the locality for 1-d configurations. In one dimension the sum in eq.(2.12) can be performed and the fermion propagator takes the form
$$\Delta^f(p) = \frac{1}{m} - \frac{2}{m^2}[\coth(m/2) - i\cot(p/2)]^{-1} + a. \tag{2.13}$$
Choosing $a = (\hat{m} - m)/m^2$, where $\hat{m} = e^m - 1$, the propagator reduces to
$$\Delta^f(p) = \left(\frac{\hat{m}}{m}\right)^2 (i\sin p + \hat{m} + \frac{1}{2}\hat{p}^2)^{-1}. \tag{2.14}$$
This corresponds to the standard nearest neighbor Wilson fermion action. Hence the perfect action is ultralocal for 1-dimensional configurations. At $m = 0$ the optimal parameter is $a = 1/2$. The same fixed point was identified before by numerically optimizing the locality in two dimensions [23]. It turns out that the perfect action is still very local even in $d = 4$. To see this we go back to coordinate space and we write the action as
$$S[\bar{\Psi}, \Psi] = \sum_{x,y} \bar{\Psi}_x[\gamma_\mu \rho_\mu(x-y) + \lambda(x-y)]\Psi_y. \tag{2.15}$$
Performing a Fourier transform of the inverse fermion propagator for the optimal value $a = (\hat{m} - m)/m^2$ we find $\rho_\mu(z)$ and $\lambda(z)$. Their values for $m = 0, 1$ and $2$ are given in table 1 and 2, and are illustrated in fig.1 and 2. For a lattice spacing of 0.3 fm (which may be realistic for a simulation with a perfect action) the value $m = 2$ roughly corresponds to the mass of the charm quark.

| $z_1$ | $z_2$ | $z_3$ | $z_4$ | $m=0$ | $m=1$ | $m=2$ |
|---|---|---|---|---|---|---|
| 1 | 0 | 0 | 0 | 0.1400 | 0.05503 | 0.01856 |
| 1 | 1 | 0 | 0 | 0.0314 | 0.01088 | 0.00314 |
| 1 | 1 | 1 | 0 | 0.0096 | 0.00302 | 0.00076 |
| 1 | 1 | 1 | 1 | 0.0033 | 0.00100 | 0.00023 |
| 1 | 2 | 0 | 0 | 0.0015 | 0.00024 | 0.00003 |
| 1 | 2 | 1 | 0 | 0.0006 | 0.00010 | 0.00001 |
| 1 | 2 | 1 | 1 | 0.0002 | 0.00003 | 0.00000 |
| 1 | 3 | 0 | 0 | 0.0001 | 0.00001 | 0.00000 |
| 2 | 0 | 0 | 0 | 0.0040 | 0.00054 | 0.00003 |
| 2 | 1 | 0 | 0 | 0.0012 | 0.00018 | 0.00002 |
| 2 | 1 | 1 | 0 | 0.0002 | 0.00004 | 0.00001 |
| 2 | 1 | 1 | 1 | -0.0002 | -0.00002 | 0.00000 |
| 2 | 2 | 0 | 0 | -0.0001 | -0.00003 | -0.00001 |
| 2 | 2 | 1 | 0 | -0.0001 | -0.00003 | 0.00000 |
| 2 | 2 | 1 | 1 | -0.0002 | -0.00002 | 0.00000 |
| 3 | 0 | 0 | 0 | 0.0004 | 0.00004 | 0.00000 |
| 3 | 1 | 0 | 0 | 0.0001 | 0.00001 | 0.00000 |

Table 1: *Some values of $\rho_1(z)$ in $d=4$ for masses $m=0,1$ and $2$. The parameter $a$ is chosen such that the action is ultralocal for 1-d configurations. Note that $\rho_1(z)$ is symmetric under permutations of $z_2$, $z_3$ and $z_4$.*

| $z_1$ | $z_2$ | $z_3$ | $z_4$ | $m=0$ | $m=1$ | $m=2$ |
|---|---|---|---|---|---|---|
| 0 | 0 | 0 | 0 | 1.8530 | 1.26893 | 0.84425 |
| 1 | 0 | 0 | 0 | -0.0616 | -0.03039 | -0.01206 |
| 1 | 1 | 0 | 0 | -0.0290 | -0.01067 | -0.00325 |
| 1 | 1 | 1 | 0 | -0.0146 | -0.00446 | -0.00111 |
| 1 | 1 | 1 | 1 | -0.0077 | -0.00205 | -0.00043 |
| 2 | 0 | 0 | 0 | 0.0010 | 0.00034 | 0.00009 |
| 2 | 1 | 0 | 0 | -0.0007 | -0.00012 | -0.00001 |
| 2 | 1 | 1 | 0 | -0.0007 | -0.00013 | -0.00002 |
| 2 | 1 | 1 | 1 | -0.0005 | -0.00008 | -0.00001 |
| 2 | 2 | 0 | 0 | 0.0004 | 0.00008 | 0.00001 |
| 2 | 2 | 1 | 0 | 0.0002 | 0.00004 | 0.00001 |
| 2 | 2 | 1 | 1 | 0.0001 | 0.00002 | 0.00000 |
| 2 | 2 | 2 | 0 | 0.0001 | 0.00002 | 0.00000 |
| 2 | 2 | 2 | 1 | 0.0001 | 0.00001 | 0.00000 |

Table 2: *Some values of $\lambda(z)$ in $d = 4$ for masses $m = 0, 1$ and $2$. Again the parameter a is chosen such that the action is ultralocal for 1-d configurations. Note that $\lambda(z)$ is symmetric under permutations of the $z_\mu$.*

Let us discuss the spectrum of the fermionic lattice theory. We investigate the 2-point correlation function of a fermionic operator with definite spatial momentum $\vec{p} \in ]-\pi, \pi]^{d-1}$,

$$\Psi(\vec{p})_{x_d} = \frac{1}{2\pi} \int_{-\pi}^{\pi} dp_d \ \Psi(p) \exp(ip_d x_d), \tag{2.16}$$

that creates a fermion at Euclidean time $x_d$ and the operator $\bar{\Psi}_{(\vec{x},0)}$ that annihilates a fermion at the point $\vec{x}$ at Euclidean time 0. The correlation function is given by

$$\begin{aligned}
\langle \bar{\Psi}_{(\vec{x},0)} \Psi(\vec{p})_{x_d} \rangle &= \frac{1}{2\pi} \int_{-\pi}^{\pi} dp_d \ \text{Tr} \Delta^f(p) \exp(ip_d x_d) \\
&= \frac{1}{2\pi} \int_{-\pi}^{\pi} dp_d \ \text{Tr} \Big[ \sum_{l \in \mathbf{Z}^d} \delta^f(p + 2\pi l) \Pi(p + 2\pi l)^2 + a \Big] \exp(ip_d x_d) \\
&= \frac{1}{2\pi} \int_{-\infty}^{\infty} dp_d \sum_{\vec{l} \in \mathbf{Z}^{d-1}} \frac{m}{(\vec{p} + 2\pi \vec{l})^2 + p_d^2 + m^2} \\
&\quad \times \prod_{j=1}^{d-1} \Big( \frac{\hat{p}_j}{p_j + 2\pi l_j} \Big)^2 \Big( \frac{\hat{p}_d}{p_d} \Big)^2 \exp(ip_d x_d) + a\delta_{x_d, 0}
\end{aligned}$$

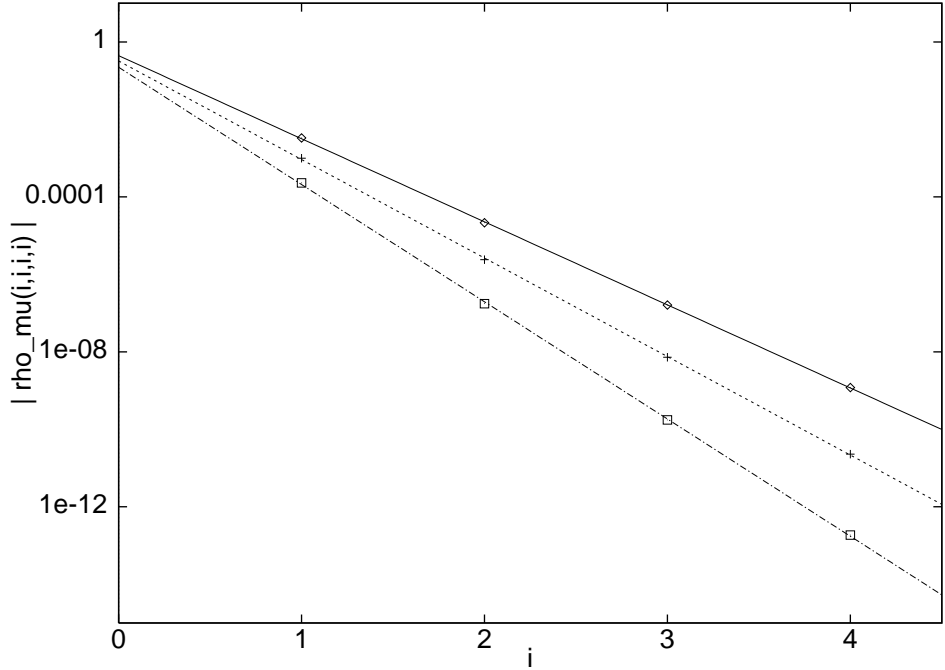

Figure 1: *Exponential decay of $|\rho_\mu(i,i,i,i)|$ (along the diagonal) for masses $m = 0$ (diamonds), $m = 1$ (crosses) and $m = 2$ (squares). The lines are least square fits of $|\rho_\mu(i,i,i,i)| \propto \exp(-c(m)i)$ with $c(0) = 4.939$, $c(1) = 5.857$ and $c(2) = 6.969$.*

$$= \sum_{\vec{l} \in \mathbf{Z}^{d-1}} C(\vec{p} + 2\pi\vec{l})\exp(-E(\vec{p}+2\pi\vec{l})x_d) + a\delta_{x_d,0}. \qquad (2.17)$$

The sum over $l_d$ has been combined with the integral of $p_d$ over $]-\pi,\pi]$ to an integral over **R**, characteristic for the continuum theory. The sum over spatial $\vec{l}$ leads to infinitely many poles of the integrand, and thus to infinitely many states that contribute an exponential to the 2-point function. The energies of these states are given by the location of the poles, $E(\vec{p}+2\pi\vec{l}) = -ip_d$, with

$$E(\vec{p}+2\pi\vec{l})^2 = -p_d^2 = (\vec{p}+2\pi\vec{l})^2 + m^2. \qquad (2.18)$$

This is exactly the energy of a particle with mass $m$ and momentum $\vec{p}+2\pi\vec{l}$, i.e. the spectrum of the lattice theory is identical with the one of the continuum — cut-off effects are completely eliminated. This is possible only because the lattice momentum $\vec{p}$, which is restricted to the Brillouin zone, combines naturally with the integer vectors $\vec{l}$ to cover the whole momentum space of the continuum theory. Therefore the continuous Poincaré symmetry is restored in the spectrum, although it is explicitly broken in the action.

The nonperturbative construction of an approximately perfect action for QCD will require some numerical work. Then it is essential to use a renormalization group transformation with a finite blocking factor. In section 6 we will define a factor $n$ transformation whose perfect action is identical with the one of blocking from the continuum at least for weak fields. To optimize the parameters of the factor $n$ transformation we want to relate them to those of blocking from the continuum. In the discrete transformation the fermionic variables $\bar{\Psi}_x, \Psi_x$ of an $n^d$ block on the fine lattice are averaged to form coarse lattice variables

$$\bar{\Psi}'_{x'} = \frac{b_n}{n^d} \sum_{x \in x'} \bar{\Psi}_x, \quad \Psi'_{x'} = \frac{b_n}{n^d} \sum_{x \in x'} \Psi_x \qquad (2.19)$$

located at the block centers $x'$ that form a lattice of spacing $n$. (By $x \in x'$ we denote that the fine lattice point $x$ belongs to the block $x'$.) The factor $b_n$ is a wave function renormalization that is necessary because on the coarse lattice we again work in lattice units (now of the coarse lattice). The renormalization group transformation with block factor $n$ takes the form

$$\begin{aligned}
\exp(-S'[\bar{\Psi}', \Psi']) &= \int \mathcal{D}\bar{\Psi}\mathcal{D}\Psi\mathcal{D}\bar{\eta}\mathcal{D}\eta \exp\left\{ -\frac{1}{(2\pi)^d} \int_B d^d p \; \bar{\Psi}(-p)\Delta^f(p)^{-1}\Psi(p) \right\} \\
&\times \exp\left\{ \left(\frac{n}{2\pi}\right)^d \int_{B'} d^d p \Big[ a_n \bar{\eta}(-p)\eta(p) \right. \\
&+ \left. [\bar{\Psi}'(-p) - b_n \sum_{l'} \bar{\Psi}(-p - \frac{2\pi l'}{n}) \Pi_n(p + \frac{2\pi l'}{n}) ]\eta(p) \right.
\end{aligned}$$

$$+ \quad \bar{\eta}(-p)[\Psi'(p) - b_n \sum_{l'} \Psi(p + \frac{2\pi l'}{n})\Pi_n(p + \frac{2\pi l'}{n})]\Big\}. \qquad (2.20)$$

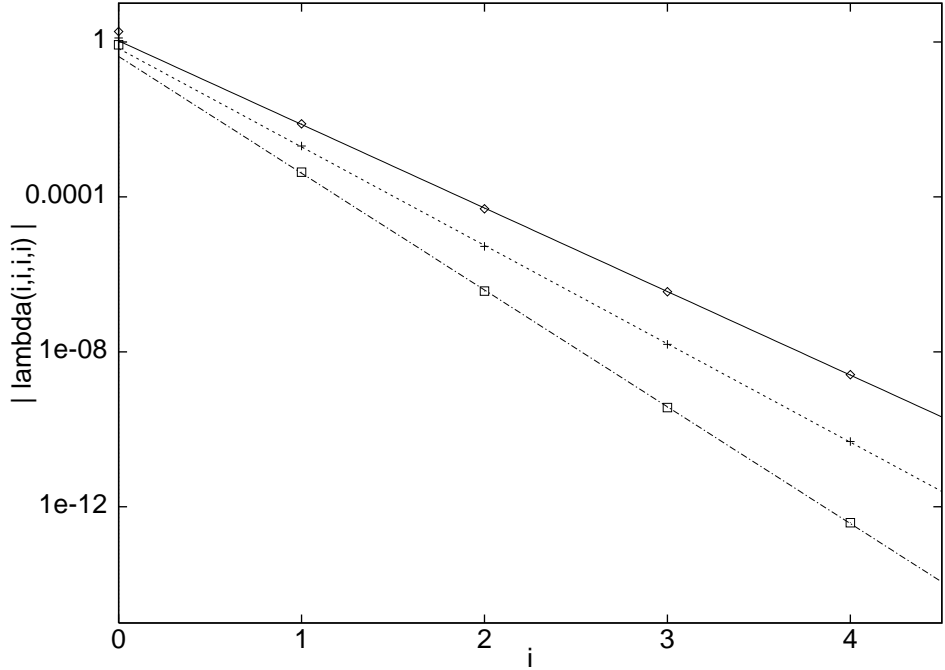

Figure 2: *Exponential decay of $|\lambda(i,i,i,i)|$ (along the diagonal) for masses $m = 0$ (diamonds), $m = 1$ (crosses) and $m = 2$ (squares). The lines are fits of $|\lambda(i,i,i,i)| \propto \exp(-c(m)i)$ excluding the point $(0,0,0,0)$. Here $c(0) = 4.967$, $c(1) = 5.850$ and $c(2) = 6.942$ are very close to the corresponding values for $|\rho_\mu(i,i,i,i)|$.*

Here $B' = ]-\pi/n, \pi/n]^d$ is the Brillouin zone of the coarse lattice. As before we have introduced a smearing parameter $a_n$. The sum over $l'$ now extends over integer vectors with components restricted to $l'_\mu \in \{1, 2, ..., n\}$ and $\Pi_n(p) = \Pi(np)/\Pi(p)$. Integrating out the fine lattice and auxiliary fields we obtain the propagator of the coarse lattice field

$$\begin{aligned}
\Delta^{f'}(p) &= \frac{b_n^2}{n^d} \sum_{l'} \Delta^f(\frac{p + 2\pi l'}{n})\Pi_n(\frac{p + 2\pi l'}{n})^2 + a_n \\
&= \frac{b_n^2}{n^d} \sum_{l'} \Big[ \sum_{l \in \mathbf{Z}^d} [i\gamma_\mu(\frac{p_\mu + 2\pi l'_\mu}{n} + 2\pi l_\mu) + m]^{-1} \Pi(\frac{p + 2\pi l'}{n} + 2\pi l)^2 + a \Big] \\
&\quad \times \Pi_n(\frac{p + 2\pi l'}{n})^2 + a_n \\
&= \frac{b_n^2}{n^d} \sum_{l' \in \mathbf{Z}^d} [i\gamma_\mu \frac{p_\mu + 2\pi l'_\mu}{n} + m]^{-1} \Pi(p + 2\pi l')^2 + \frac{b_n^2}{n^d} \sum_{l'} a\Pi_n(\frac{p + 2\pi l'}{n})^2 + a_n \\
&= \frac{b_n^2}{n^{d-1}} \sum_{l' \in \mathbf{Z}^d} [i\gamma_\mu(p_\mu + 2\pi l'_\mu) + nm]^{-1} \Pi(p + 2\pi l')^2 + \frac{b_n^2}{n^d}a + a_n. \qquad (2.21)
\end{aligned}$$

This is indeed the perfect propagator for blocking from the continuum provided that $b_n = n^{(d-1)/2}$ and $a/n + a_n = a$. The value of $b_n$ is understood on dimensional grounds because the dimension of a fermion field is $(d-1)/2$. The mass is a factor $n$ larger because it is now measured in units of the coarse lattice. One obtains [24]

$$a_n = a\left(1 - \frac{1}{n}\right). \qquad (2.22)$$

This means that e.g. the optimal value $a_2$ for a factor 2 renormalization group transformation is half the value of $a$ that is optimal for blocking from the continuum. Also $a_\infty = a$ because blocking from the continuum is equivalent to blocking with an infinite blocking factor.

## 3  Blocking free gluons from the continuum

In this section we apply the technique of blocking from the continuum to free gluons. As long as interactions are switched off the gluons behave like decoupled Abelian gauge fields. Therefore we suppress color indices and denote the gluon field in the continuum by $a_\mu$. In analogy to the fermions we want to average the continuum gauge field to obtain a blocked lattice field. Lattice gauge fields naturally live on the links connecting neighboring lattice points. The lattice field is therefore obtained by an integration over the two hypercubes centered at these lattice points. We define a noncompact lattice gauge field

$$A_{\mu,x} = \int_{c_{x-\hat{\mu}/2}} d^d y \, (1 + y_\mu - x_\mu) a_\mu(y) + \int_{c_{x+\hat{\mu}/2}} d^d y \, (1 - y_\mu + x_\mu) a_\mu(y), \quad (3.1)$$

where $x$ now refers to the center of the link, i.e. $x_\mu$ is a half-integer. It is remarkable that this construction is gauge covariant. When one performs a gauge transformation

$$^\varphi a_\mu(x) = a_\mu(x) - \partial_\mu \varphi(x) \quad (3.2)$$

in the continuum, this induces a lattice gauge transformation

$$^\Phi A_{\mu,x} = A_{\mu,x} - \Phi_{x+\hat{\mu}/2} + \Phi_{x-\hat{\mu}/2}. \quad (3.3)$$

The lattice gauge transformation is simply given by the integrated continuum transformation, i.e.

$$\Phi_x = \int_{c_x} d^d y \, \varphi(y). \quad (3.4)$$

Note that it is sufficient to consider Abelian gauge transformations because we have not yet switched on interactions. A $\delta$-function renormalization group transformation for the gauge field takes the form

$$\exp(-S[A]) = \int \mathcal{D}a \, \exp(-s[a])$$
$$\times \prod_{x,\mu} \delta(A_{\mu,x} - \int_{c_{x-\hat{\mu}/2}} d^d y \, (1 + y_\mu - x_\mu) a_\mu(y) - \int_{c_{x+\hat{\mu}/2}} d^d y \, (1 - y_\mu + x_\mu) a_\mu(y)), \quad (3.5)$$

where $s[a]$ is the continuum gauge action. In momentum space the lattice gauge field is given by

$$A_\mu(p) = \sum_{l \in \mathbf{Z}^d} a_\mu(p + 2\pi l) \Pi_\mu(p + 2\pi l)(-1)^{l_\mu}, \quad \Pi_\mu(p) = \frac{\hat{p}_\mu}{p_\mu} \Pi(p). \quad (3.6)$$

Because the link centers have a half-integer component $x_\mu$, the gauge field is antiperiodic over the Brillouin zone in the $\mu$-direction. For technical reasons we work in the Landau gauge $\sum_\mu p_\mu a_\mu(p) = 0$, i.e. the measure $\mathcal{D}a$ contains $\delta$-functions $\delta(\sum_\mu p_\mu a_\mu(p))$. Then the continuum action takes the form

$$s[a] = \frac{1}{(2\pi)^d} \int d^d p \, \frac{1}{2} a_\mu(-p) \delta^g(p)^{-1} a_\mu(p), \quad \delta^g(p) = p^{-2}, \quad (3.7)$$

where $\delta^g(p)$ is the free gluon propagator in the Landau gauge. As in the fermionic case we want to optimize the renormalization group transformation such that the fixed point action becomes as local as possible. Therefore we generalize the renormalization group transformation of eq.(3.5) by smearing the $\delta$-function. For this purpose we introduce an auxiliary lattice vector field $D_\mu$, analogous to the auxiliary fermion field $\eta$. Also for $D_\mu$ we introduce a mass term $\frac{1}{2}\alpha D_\mu D_\mu$ to smear the $\delta$-function. This alone would, however, not lead to an optimally local fixed point. Therefore we introduce in addition a kinetic term $\frac{1}{2}\gamma \hat{p}_\mu^2 D_\mu D_\mu$ for the auxiliary field. This will allow us to optimize the parameters in the renormalization group transformation such that the fixed point action becomes ultralocal even for 2-d configurations. The introduction of a kinetic term for an auxiliary field was used before for staggered fermions [8, 25]. It is easy to show that the partition function remains invariant when one smears the $\delta$-function in this way. We write the renormalization group transformation as

$$
\begin{aligned}
\exp(-S[A]) &= \int \mathcal{D}a\mathcal{D}D \exp\Big\{ -\frac{1}{(2\pi)^d} \int d^d p\, \frac{1}{2} a_\mu(-p)\delta^g(p)^{-1} a_\mu(p) \Big\} \\
&\times \exp\Big\{ -\frac{1}{(2\pi)^d} \int_B d^d p \Big[\frac{1}{2} D_\mu(-p)(\alpha + \gamma \hat{p}_\mu^2) D_\mu(p) \\
&+ iD_\mu(-p)[A_\mu(p) - \sum_{l \in \mathbf{Z}^d} a_\mu(p+2\pi l)\Pi_\mu(p+2\pi l)(-1)^{l_\mu}]\Big]\Big\} \\
&= \int \mathcal{D}a\mathcal{D}D \exp\Big\{ -\frac{1}{(2\pi)^d} \int d^d p\Big[\frac{1}{2} a_\mu(-p)\delta^g(p)^{-1} a_\mu(p) \\
&- iD_\mu(-p)a_\mu(p)\Pi_\mu(p)\Big]\Big\} \\
&\times \exp\Big\{ -\frac{1}{(2\pi)^d} \int_B d^d p\Big[\frac{1}{2} D_\mu(-p)(\alpha+\gamma\hat{p}_\mu^2)D_\mu(p) \\
&+ iD_\mu(-p)A_\mu(p)\Big]\Big\}. \qquad (3.8)
\end{aligned}
$$

Performing the Gaussian integral over $a_\mu$ is equivalent to solving a classical equation of motion which yields

$$a_{\mu c}(p) = i\delta^g(p)\Pi_\mu(p)D_\mu(p). \qquad (3.9)$$

Since the continuum gauge field obeys the Landau gauge condition $\sum_\mu p_\mu a_\mu(p) = 0$ we find

$$\sum_\mu p_\mu \delta^g(p)\Pi_\mu(p)D_\mu(p) = 0. \qquad (3.10)$$

Using $p_\mu \Pi_\mu(p) = \hat{p}_\mu \Pi(p)$ one obtains $\sum_\mu \hat{p}_\mu D_\mu(p) = 0$, i.e. the auxiliary field $D_\mu$ obeys the standard lattice Landau gauge condition. Now we take the solution of eq.(3.9) and insert it in eq.(3.8). This is equivalent to integrating out the continuum gluons and yields

$$
\begin{aligned}
\exp(-S[A]) &= \int \mathcal{D}D \exp\Big\{ -\frac{1}{(2\pi)^d}\int_B d^d p\Big[iD_\mu(-p)A_\mu(p) \\
&+ \frac{1}{2}D_\mu(-p)[\sum_{l \in \mathbf{Z}^d}\delta^g(p+2\pi l)\Pi_\mu(p+2\pi l)^2 + \alpha + \gamma\hat{p}_\mu^2]D_\mu(p)\Big]\Big\}.
\end{aligned}
$$
(3.11)

Finally we integrate out the auxiliary vector field, i.e. we again solve a classical equation of motion, which now takes the form

$$D_{\mu c}(p) = -i\Delta_\mu^g(p)^{-1} A_\mu(p), \tag{3.12}$$

with

$$\Delta_\mu^g(p) = \sum_{l \in \mathbf{Z}^d} \delta^g(p + 2\pi l)\Pi_\mu(p + 2\pi l)^2 + \alpha + \gamma \hat{p}_\mu^2. \tag{3.13}$$

Combining eq.(3.12) with eq.(3.9) we obtain

$$a_{\mu c}(p) = \delta^g(p)\Pi_\mu(p)\Delta_\mu^g(p)^{-1} A_\mu(p). \tag{3.14}$$

This equation will be useful in sections 4 and 5 when we construct composite operators and determine the vertex functions. The auxiliary vector field is in the standard lattice Landau gauge and hence one obtains

$$\sum_\mu \hat{p}_\mu \Delta_\mu^g(p)^{-1} A_\mu(p) = 0. \tag{3.15}$$

We call this the 'fixed point lattice Landau gauge' condition, because it is the one that arises from blocking gauge fields that obey the Landau gauge condition in the continuum. Then the measure $\mathcal{D}A$ contains $\delta$-functions $\delta(\sum_\mu \hat{p}_\mu \Delta_\mu^g(p)^{-1} A_\mu(p))$. Inserting eq.(3.12) in eq.(3.11) one arrives at the fixed point action

$$\begin{aligned} S[A] &= \frac{1}{(2\pi)^d} \int_B d^d p\, \frac{1}{2} A_\mu(-p)\Delta_\mu^g(p)^{-1} A_\mu(p), \\ \Delta_\mu^g(p) &= \sum_{l \in \mathbf{Z}^d} (p + 2\pi l)^{-2} \Pi_\mu(p + 2\pi l)^2 + \alpha + \gamma \hat{p}_\mu^2. \end{aligned} \tag{3.16}$$

Here $\Delta_\mu^g(p)$ is the lattice gluon propagator in the fixed point lattice Landau gauge. Of course, we would like to express the fixed point action in a gauge invariant way, i.e. we would like to go back to an unspecified gauge. Suppose the gauge field in an arbitrary gauge is given by $^\Phi A_\mu$. Then $^\Phi A_\mu$ is related to the gauge field $A_\mu$ in the fixed point lattice Landau gauge by a gauge transformation $\Phi$, i.e. $^\Phi A_\mu(p) = A_\mu(p) - i\hat{p}_\mu \Phi(p)$. The gauge condition (3.15) is an equation for $\Phi$ which yields

$$\Phi(p) = i\frac{\sum_\nu \hat{p}_\nu \Delta_\nu^g(p)^{-1}\, {}^\Phi A_\nu(p)}{\sum_\rho \hat{p}_\rho^2 \Delta_\rho^g(p)^{-1}}, \quad A_\mu(p) = {}^\Phi A_\mu(p) - \hat{p}_\mu \frac{\sum_\nu \hat{p}_\nu \Delta_\nu^g(p)^{-1}\, {}^\Phi A_\nu(p)}{\sum_\rho \hat{p}_\rho^2 \Delta_\rho^g(p)^{-1}}. \tag{3.17}$$

Inserting $A_\mu$ into eq.(3.16) and dropping the index $\Phi$ on $^\Phi A_\mu$, one obtains the fixed point action in a gauge invariant form

$$\begin{aligned} S[A] &= \frac{1}{(2\pi)^d} \int_B d^d p\, \frac{1}{2} A_\mu(-p)\Delta_{\mu\nu}^g(p)^{-1} A_\nu(p), \\ \Delta_{\mu\nu}^g(p)^{-1} &= \Delta_\mu^g(p)^{-1}\delta_{\mu\nu} - \frac{\hat{p}_\mu \Delta_\mu^g(p)^{-1}\Delta_\nu^g(p)^{-1}\hat{p}_\nu}{\sum_\rho \hat{p}_\rho^2 \Delta_\rho^g(p)^{-1}}. \end{aligned} \tag{3.18}$$

Again we are looking for the optimally local fixed point action. It turns out that $\alpha$ and $\gamma$ can be optimized such that the action becomes an ultralocal single plaquette action for 2-d configurations. To see this it is advantageous to work with field strength variables $F$ that live on the plaquettes. In the standard lattice Landau gauge the gauge potential in $d = 2$ is related to the plaquette variables by

$$A_\mu(p) = i\hat{p}^{-2}\varepsilon_{\mu\nu}\hat{p}_\nu F(p). \tag{3.19}$$

Inserting this into eq.(3.18), which is valid in any gauge — and hence also in the standard lattice Landau gauge we are using now — one finds after some manipulations

$$S[F] = \frac{1}{(2\pi)^2}\int_B d^2p\, \frac{1}{2}F(-p)\rho(p)F(p), \quad \rho(p)^{-1} = \hat{p}_1^2\Delta_2^g(p) + \hat{p}_2^2\Delta_1^g(p). \tag{3.20}$$

Using eq.(3.13) one obtains

$$\begin{aligned}\rho(p)^{-1} &= \sum_{l\in\mathbf{Z}^2}\delta^g(p+2\pi l)[\hat{p}_1^2\Pi_2(p+2\pi l)^2 + \hat{p}_2^2\Pi_1(p+2\pi l)^2] + \alpha(\hat{p}_1^2+\hat{p}_2^2) + 2\gamma\hat{p}_1^2\hat{p}_2^2 \\ &= \sum_{l\in\mathbf{Z}^2}\Pi(p+2\pi l)^4 + \alpha(\hat{p}_1^2+\hat{p}_2^2) + 2\gamma\hat{p}_1^2\hat{p}_2^2 \\ &= (1-\frac{1}{6}\hat{p}_1^2)(1-\frac{1}{6}\hat{p}_2^2) + \alpha(\hat{p}_1^2+\hat{p}_2^2) + 2\gamma\hat{p}_1^2\hat{p}_2^2.\end{aligned} \tag{3.21}$$

For

$$\alpha = \frac{1}{6}, \quad \gamma = -\frac{1}{72}, \tag{3.22}$$

one arrives at $\rho(p) = 1$. This corresponds to the standard plaquette action which is ultralocal. Hence at least in $d = 2$ the optimal choice for the smearing term is $\frac{1}{12}D_\mu(1-\frac{1}{12}\hat{p}_\mu^2)D_\mu$. We use the same parameters in $d = 4$ and we write the fixed point action in coordinate space as

$$S[A] = \sum_{x,y}\frac{1}{2}A_{\mu,x}\rho_{\mu\nu}(x-y)A_{\nu,y}. \tag{3.23}$$

Performing a Fourier transformation of eq.(3.18) one finds the values of $\rho_{\mu\nu}(z)$ that are given in table 3. As shown in fig.3 even in $d = 4$ the fixed point action is very local (although not ultralocal). In [14] we will systematically vary $\alpha$ and $\gamma$ to optimize locality in $d = 4$. There a parametrization of the action in terms of closed loops will be given as well.

Also for the gluons we would like to see that the spectrum of the lattice theory is perfect. We proceed in complete analogy to the quarks. For simplicity we work in the fixed point lattice Landau gauge. We construct an operator

$$A_i(\vec{p})_{x_d} = \frac{1}{2\pi}\int_{-\pi}^{\pi}dp_d\, A_i(p)\exp(ip_dx_d), \tag{3.24}$$

| $z_1$ | $z_2$ | $z_3$ | $z_4$ | $\rho_{11}(z)$ | $z_1$ | $z_2$ | $z_3$ | $z_4$ | $\rho_{12}(z)$ |
|---|---|---|---|---|---|---|---|---|---|
| 0 | 0 | 0 | 0 | 4.1891 | 0.5 | 0.5 | 0 | 0 | -0.7120 |
| 0 | 1 | 0 | 0 | -0.4437 | 0.5 | 0.5 | 1 | 0 | -0.0640 |
| 0 | 1 | 1 | 0 | -0.1050 | 0.5 | 0.5 | 1 | 1 | -0.0106 |
| 0 | 1 | 1 | 1 | -0.0324 | 0.5 | 0.5 | 2 | 0 | 0.0023 |
| 0 | 2 | 0 | 0 | 0.0030 | 0.5 | 0.5 | 2 | 1 | 0.0001 |
| 0 | 2 | 1 | 0 | -0.0008 | 1.5 | 0.5 | 0 | 0 | 0.0143 |
| 1 | 0 | 0 | 0 | -0.0831 | 1.5 | 0.5 | 1 | 0 | -0.0028 |
| 1 | 1 | 0 | 0 | 0.0267 | 1.5 | 0.5 | 1 | 1 | -0.0009 |
| 1 | 1 | 1 | 0 | -0.0038 | 1.5 | 1.5 | 0 | 0 | -0.0024 |
| 1 | 1 | 1 | 1 | -0.0033 | 1.5 | 1.5 | 1 | 0 | 0.0002 |
| 1 | 2 | 0 | 0 | -0.0024 | 1.5 | 1.5 | 1 | 1 | 0.0002 |
| 2 | 0 | 0 | 0 | 0.0027 | 1.5 | 1.5 | 2 | 0 | 0.0001 |
| 2 | 1 | 0 | 0 | -0.0010 | 2.5 | 0.5 | 0 | 0 | -0.0005 |
| 2 | 1 | 1 | 0 | 0.0003 | 2.5 | 0.5 | 1 | 0 | 0.0001 |
| 3 | 0 | 0 | 0 | -0.0001 | 2.5 | 1.5 | 0 | 0 | 0.0001 |

Table 3: *Some values of $\rho_{11}(z)$ and $\rho_{12}(z)$ in $d = 4$. The parameters $\alpha$ and $\gamma$ are chosen such that the action is ultralocal for 2-d configurations. Note that $\rho_{11}(z)$ is symmetric under permutations of $z_2$, $z_3$ and $z_4$ while $\rho_{12}(z)$ is symmetric under exchange of $z_1$ and $z_2$ as well as $z_3$ and $z_4$.*

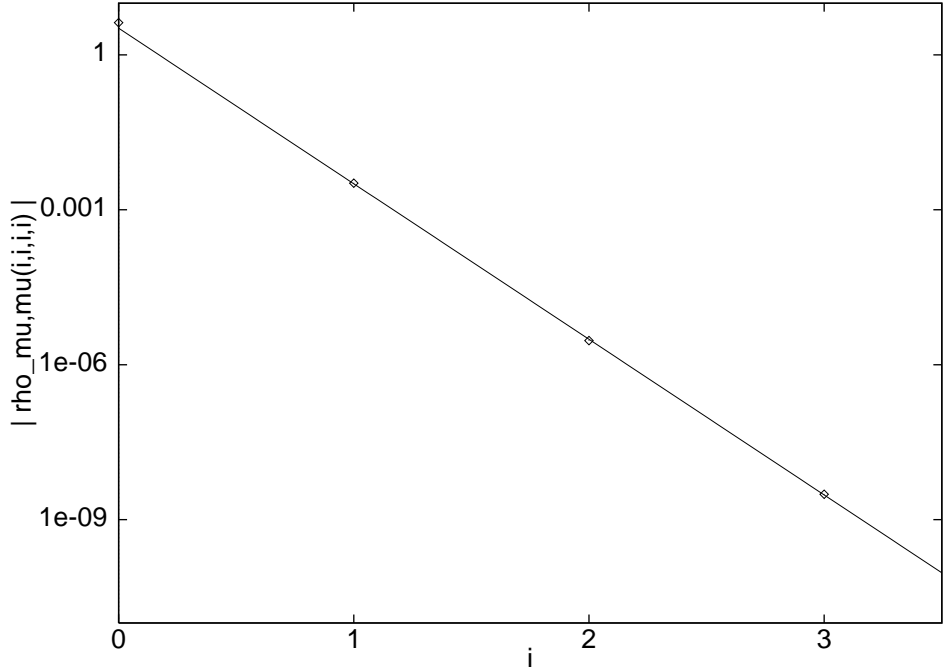

Figure 3: *Exponential decay of $|\rho_{\mu\mu}(i,i,i,i)|$ for the free gluon (along the diagonal). The straight line is an exponential fit $|\rho_{\mu\mu}(i,i,i,i)| \propto \exp(-ci)$ with $c = 6.937$ excluding the point $(0,0,0,0)$.*

and we consider the 2-point correlation function

$$
\begin{aligned}
\langle A_{i,(\vec{x},0)} A_i(\vec{p})_{x_d} \rangle &= \frac{1}{2\pi} \int_{-\pi}^{\pi} dp_d \; \Delta_i^g(p) \exp(ip_d x_d) \\
&= \frac{1}{2\pi} \int_{-\pi}^{\pi} dp_d \; [\sum_{l \in \mathbf{Z}^d} (p + 2\pi l)^{-2} \Pi_i (p + 2\pi l)^2 + \alpha + \gamma \hat{p}_i^2] \exp(ip_d x_d) \\
&= \frac{1}{2\pi} \int_{-\infty}^{\infty} dp_d \sum_{\vec{l} \in \mathbf{Z}^{d-1}} \frac{1}{(\vec{p} + 2\pi \vec{l})^2 + p_d^2} \Big(\frac{\hat{p}_i}{p_i + 2\pi l_i}\Big)^2 \prod_{j=1}^{d-1} \Big(\frac{\hat{p}_j}{p_j + 2\pi l_j}\Big)^2 \\
&\quad \times \Big(\frac{\hat{p}_d}{p_d}\Big)^2 \exp(ip_d x_d) + (\alpha + \gamma \hat{p}_i^2) \delta_{x_d, 0} \\
&= \sum_{\vec{l} \in \mathbf{Z}^{d-1}} C(\vec{p} + 2\pi \vec{l}) \exp(-E(\vec{p} + 2\pi \vec{l}) x_d) + (\alpha + \gamma \hat{p}_i^2) \delta_{x_d, 0}. \quad (3.25)
\end{aligned}
$$

Again the energies of the lattice theory are given by the locations of the poles. Now

$$E(\vec{p} + 2\pi \vec{l})^2 = -p_d^2 = (\vec{p} + 2\pi \vec{l})^2, \quad (3.26)$$

which is the energy of a massless gluon with momentum $\vec{p} + 2\pi \vec{l}$, i.e. again the spectrum is identical with the one of the continuum theory.

As for the quarks we also want to discuss renormalization group transformations with a finite blocking factor $n$. In this transformation we construct blocked link

variables
$$A'_{\mu,x'+n\hat{\mu}/2} = \frac{\beta_n}{n^d} \sum_{x \in x'} \frac{1}{n} \sum_{i=0}^{n-1} A_{\mu,x+\hat{\mu}/2+i\hat{\mu}}. \qquad (3.27)$$

We average over all straight line connections between corresponding sites in adjacent blocks. The factor $\beta_n$ is a wave function renormalization. In momentum space the renormalization group transformation takes the form

$$\begin{aligned}
\exp(-S'[A']) &= \int \mathcal{D}A\mathcal{D}D \exp\left\{ -\frac{1}{(2\pi)^d} \int_B d^d p \, \frac{1}{2} A_\mu(-p) \Delta^g_\mu(p)^{-1} A_\mu(p) \right\} \\
&\times \exp\left\{ -\left(\frac{n}{2\pi}\right)^d \int_{B'} d^d p \left[ \frac{1}{2} D_\mu(-p)(\alpha_n + \gamma_n \hat{p}^2_{n\mu}) D_\mu(p) \right.\right. \\
&\quad + iD_\mu(-p)[A'_\mu(p) - \beta_n \sum_{l'} A_\mu(p + \frac{2\pi l'}{n}) \Pi_{n\mu}(p + \frac{2\pi l'}{n})(-1)^{l'_\mu}] \Big] \Big\}.
\end{aligned} \qquad (3.28)$$

The smearing parameters are now given by $\alpha_n$ and $\gamma_n$. Also $\Pi_{n\mu}(p) = \Pi_\mu(np)/\Pi_\mu(p)$ and the summation extends over integer vectors with components $l'_\mu \in \{1, 2, ..., n\}$. Integrating out the fine lattice fields one obtains the coarse gluon propagator

$$\begin{aligned}
\Delta^{g'}_\mu(p) &= \frac{\beta_n^2}{n^d} \sum_{l'} \Delta^g\left(\frac{p+2\pi l'}{n}\right) \Pi_{n\mu}\left(\frac{p+2\pi l'}{n}\right)^2 + \alpha_n + \gamma_n \hat{p}^2_\mu \\
&= \frac{\beta_n^2}{n^d} \sum_{l'} \Big[ \sum_{l \in \mathbb{Z}^d} \left(\frac{p+2\pi l'}{n} + 2\pi l\right)^{-2} \Pi_\mu\left(\frac{p+2\pi l'}{n} + 2\pi l\right)^2 \\
&\quad + \alpha + 4\gamma \sin^2\left(\frac{p_\mu + 2\pi l'_\mu}{2n}\right) \Big] \Pi_{n\mu}\left(\frac{p+2\pi l'}{n}\right)^2 + \alpha_n + \gamma_n \hat{p}^2_\mu \\
&= \frac{\beta_n^2}{n^d} \sum_{l' \in \mathbb{Z}^d} \left(\frac{p_\mu + 2\pi l'_\mu}{n}\right)^{-2} \Pi_\mu(p + 2\pi l')^2 \\
&\quad + \frac{\beta_n^2}{n^d} \sum_{l'} \Big[\alpha + 4\gamma \sin^2\left(\frac{p_\mu + 2\pi l'_\mu}{2n}\right)\Big] \Pi_{n\mu}\left(\frac{p+2\pi l'}{n}\right)^2 + \alpha_n + \gamma_n \hat{p}^2_\mu \\
&= \frac{\beta_n^2}{n^{d-2}} \sum_{l' \in \mathbb{Z}^d} (p_\mu + 2\pi l'_\mu)^{-2} \Pi_\mu(p + 2\pi l')^2 \\
&\quad + \frac{\beta_n^2}{n^d}\Big[\alpha(1 - \frac{n^2-1}{6n^2}\hat{p}^2_\mu) + \gamma \frac{1}{n^2}\hat{p}^2_\mu\Big] + \alpha_n + \gamma_n \hat{p}^2_\mu.
\end{aligned} \qquad (3.29)$$

This is the perfect propagator for blocking from the continuum provided that $\beta_n = n^{(d-2)/2}$, $\alpha/n^2 + \alpha_n = \alpha$ and $\gamma/n^4 + \gamma_n - \alpha(n^2-1)/6n^4 = \gamma$. The value of $\beta_n$ is consistent with dimensional considerations because the dimension of a gauge field is $(d-2)/2$. One also obtains

$$\alpha_n = \alpha\left(1 - \frac{1}{n^2}\right), \quad \gamma_n = \gamma\left(1 - \frac{1}{n^4}\right) + \frac{\alpha}{6n^2}\left(1 - \frac{1}{n^2}\right). \qquad (3.30)$$

In particular $\alpha_\infty = \alpha$ and $\gamma_\infty = \gamma$ because blocking from the continuum corresponds to an infinite blocking factor. Using the optimized values $\alpha = 1/6$ and $\gamma = -1/72$ obtained in eq.(3.22) for blocking from the continuum one finds for general block factor $n$

$$\alpha_n = \frac{1}{6}\left(1 - \frac{1}{n^2}\right), \ \gamma_n = -\frac{1}{72}\left(1 - \frac{1}{n^2}\right)^2. \tag{3.31}$$

In particular for block factor 2 the values are $\alpha_2 = 1/8$ and $\gamma_2 = -1/128$.

# 4 Classically perfect fields, composite operators and currents

In this section we construct classically perfect expressions for quark and gluon fields, which we use to construct lattice versions of quark-gluon composite operators of the continuum theory. These operators may serve as a first approximation to quantum perfect operators that one would like to use e.g. in numerical simulations of hadronic matrix elements. In particular we construct the Polyakov loop as well as current operators. In general, the correlation functions of the classically perfect fields are not completely free of cut-off effects. However, as we show explicitly for the quark 2-point function and for the correlation function of two Polyakov loops, the cut-off effects are exponentially suppressed. We also construct other lattice currents which have an exact interpretation in terms of continuum physics. They are useful when one wants to discuss Ward identities or investigate anomalies in the perfect action approach. For example, in [22] it has been shown that the exact lattice axial current has the correct axial anomaly in the 2-d lattice Schwinger model, whereas the naive Noether current is conserved.

Let us consider the continuum quark field $\psi$. As it was observed before in scalar field theory [10], at the classical level the continuum field can be reconstructed from the lattice field $\Psi$. This was achieved in eq.(2.11) were we found

$$\bar{\psi}_c(-p) = \bar{\Psi}(-p)\Delta^f(p)^{-1}\Pi(p)\delta^f(p), \ \psi_c(p) = \delta^f(p)\Pi(p)\Delta^f(p)^{-1}\Psi(p). \tag{4.1}$$

The correlation function $\langle\bar{\psi}_c(x)\psi_c(y)\rangle$ is invariant only against simultaneous translations of $x$ and $y$ by lattice vectors, although $x$ and $y$ may be arbitrary points in the continuum. To restore full continuum translation invariance we consider $\int_{c_0} d^d z \ \langle\bar{\psi}_c(x+z)\psi_c(y+z)\rangle$, i.e. we average the correlation function over a hypercube $c_0$ centered at the origin. Note that the displacement between the quark fields is fixed. In momentum space this corresponds to

$$\frac{1}{V}\langle\bar{\psi}_c(-p)\psi_c(p)\rangle = \delta^f(p)^2\Pi(p)^2\Delta^f(p)^{-1}. \tag{4.2}$$

Here $V$ is the space-time volume. Obviously the result differs from the correct continuum expression $\frac{1}{V}\langle\bar{\psi}(-p)\psi(p)\rangle = \delta^f(p)$. The same is true in scalar field theory.

Therefore the 2-point function of the classically perfect field is *not* perfect — it is different from the one of the continuum theory, and hence it is not free of cut-off effects. Still the classically perfect field is an interesting object, because — as we will see now — the cut-off effects in its 2-point function are exponentially suppressed at large distance. From eq.(4.2) we infer

$$\langle \bar{\psi}_c(-p)\psi_c(p)\rangle = \langle \bar{\psi}(-p)\psi(p)\rangle \; \delta^f(p)\Pi(p)^2\Delta^f(p)^{-1}. \tag{4.3}$$

In coordinate space this implies that $\int_{c_0} d^d z \langle \bar{\psi}_c(x+z)\psi_c(y+z)\rangle$ is a convolution of the continuum 2-point function $\langle \bar{\psi}(x)\psi(y)\rangle$ with the Fourier transform of $\delta^f(p)\Pi(p)^2\Delta^f(p)^{-1}$ which is defined at any point in the continuum. We want to show that this function decays exponentially. For this purpose we restrict it to the lattice points. In momentum space this corresponds to $p \in B$ and a summation over integer vectors

$$\sum_{l \in \mathbf{Z}^d} \delta^f(p+2\pi l)\Pi(p+2\pi l)^2 \Delta^f(p+2\pi l)^{-1} = [\Delta^f(p)-a]\Delta^f(p)^{-1} = 1-a\Delta^f(p)^{-1}. \tag{4.4}$$

Here we have used eq.(2.10) as well as the periodicity of $\Delta^f(p)$. In coordinate space the result corresponds to $\delta_{x,y} - a[\gamma_\mu \rho_\mu(x-y) + \lambda(x-y)]$. As we have seen in section 2 this function does indeed decay exponentially. Hence the 2-point function of the classically perfect field differs from the correct 2-point function of the continuum theory only by exponentially small cut-off effects. The same was concluded in the first paper of ref.[10] for scalar field theory.

Of course we can also construct the classically perfect gluon field. In fact, this has already been done in eq.(3.14) where we found

$$a_{\mu c} = \delta^g(p)\Pi_\mu(p)\Delta^g_\mu(p)^{-1} A_\mu(p). \tag{4.5}$$

Note that in this equation $a_{\mu c}$ is in the Landau gauge and $A_\mu$ is in the fixed point lattice Landau gauge. Using eqs.(4.1) and (4.5) one can construct a classical lattice version of any quark-gluon composite operator $\mathcal{O}[\bar{\psi},\psi,a]$ of the continuum theory simply by identifying

$$\mathcal{O}_c[\bar{\Psi},\Psi,A] = \mathcal{O}[\bar{\psi}_c,\psi_c,a_c]. \tag{4.6}$$

The correlation functions of the composite operators will not be perfect. However, as for the 2-point function of the classically perfect quark field one expects only exponentially small cut-off effects. This may be useful e.g. in numerical simulations of hadronic matrix elements. As an example of a composite operator we consider the Polyakov loop

$$\phi(\vec{x}) = \int dx_d \; a_d(\vec{x},x_d). \tag{4.7}$$

Here we restrict ourselves to the leading order in the gauge coupling. Then path ordering problems do not occur. In momentum space we have $\phi(\vec{p}) = a_d(\vec{p},0)$. The Polyakov loop expressed in terms of lattice variables is then given by

$$\phi_c(\vec{p}) = \delta^g(\vec{p},0)\Pi_d(\vec{p},0)\Delta^g_d(\vec{p},0)^{-1} A_d(\vec{p},0). \tag{4.8}$$

In general the classically perfect Polyakov loop is defined at any point in the continuum coordinate space. If we restrict ourselves to lattice sites $\vec{x}$ we can write

$$\Phi_{c,\vec{x}} = \phi_c(\vec{x}) = \sum_{\vec{y}} \omega(\vec{x} - \vec{y})\Phi_{\vec{y}}, \quad \Phi_{\vec{y}} = \sum_{y_d} A_{d,(\vec{y},y_d)}, \tag{4.9}$$

where $\Phi_{\vec{y}}$ with $\vec{y} \in \mathbf{Z}^{d-1}$ is the standard lattice Polyakov loop. In momentum space the above kernel is given by

$$\omega(\vec{p}) = \sum_{\vec{l} \in \mathbf{Z}^{d-1}} (\vec{p} + 2\pi\vec{l})^{-2} \Pi_d(\vec{p} + 2\pi\vec{l}, 0) \Delta_d^g(\vec{p}, 0)^{-1}, \tag{4.10}$$

where

$$\begin{aligned}
\Delta_d^g(\vec{p}, 0) &= \sum_{l \in \mathbf{Z}^d} [(\vec{p} + 2\pi\vec{l})^2 + (2\pi l_d)^2]^{-1} \prod_{i=1}^{d-1} \left(\frac{\hat{p}_i}{p_i + 2\pi l_i}\right)^2 \left(\frac{\sin(\pi l_d)}{\pi l_d}\right)^4 + \alpha \\
&= \sum_{\vec{l} \in \mathbf{Z}^{d-1}} (\vec{p} + 2\pi\vec{l})^{-2} \prod_{i=1}^{d-1} \left(\frac{\hat{p}_i}{p_i + 2\pi l_i}\right)^2 + \alpha.
\end{aligned} \tag{4.11}$$

First we consider $d = 2$ and find

$$\Delta_2^g(p_1, 0) = \sum_{l_1 \in \mathbf{Z}} \frac{\hat{p}_1^2}{(p_1 + 2\pi l_1)^4} + \alpha = \frac{1}{\hat{p}_1^2} - \frac{1}{6} + \alpha, \tag{4.12}$$

i.e. for the value $\alpha = 1/6$, which yields an ultralocal fixed point action in $d = 2$, one obtains $\Delta_2^g(p_1, 0)^{-1} = \hat{p}_1^2$. Inserting this in eq.(4.10) one finds $\omega(0) = 3/4$ and $\omega(\pm 1) = 1/8$. At all larger distances $\omega(z)$ vanishes. Hence the classically perfect Polyakov loop in $d = 2$ is ultralocal and is given by

$$\Phi_{c,x} = \frac{1}{8}(\Phi_{x+1} + 6\Phi_x + \Phi_{x-1}) \tag{4.13}$$

Some values of $\omega(\vec{z})$ for $d = 4$ are given in table 4 and in fig.4. Again the classically perfect Polyakov loop is very local (but no longer ultralocal).

Next we consider the correlation function of two Polyakov loops. It takes the form

$$\frac{1}{V}\langle\phi_c(-\vec{p})\phi_c(\vec{p})\rangle = \delta^g(\vec{p}, 0)^2 \Pi_d(\vec{p}, 0)^2 \Delta_d^g(\vec{p}, 0)^{-1}. \tag{4.14}$$

Again this is different from the continuum result $\frac{1}{V}\langle\phi(-\vec{p})\phi(\vec{p})\rangle = \delta^g(\vec{p}, 0)$. As for the quark 2-point function we like to show that the cut-off effects are exponentially small. In coordinate space the left hand side of eq.(4.14) corresponds to $\int_{f_0} d^{d-1}z \langle\phi_c(\vec{x}+\vec{z})\phi_c(\vec{y}+\vec{z})\rangle$, where $f_0$ is a $(d-1)$-dimensional cube centered at the origin. The correlation function averaged over $f_0$ is invariant against arbitrary continuum translations. It is a convolution of the correct continuum result $\langle\phi(\vec{x})\phi(\vec{y})\rangle$

| $z_1$ | $z_2$ | $z_3$ | $\omega(\vec{z})$ |
|---|---|---|---|
| 0 | 0 | 0 | 0.48709 |
| 1 | 0 | 0 | 0.05188 |
| 1 | 1 | 0 | 0.01354 |
| 1 | 1 | 1 | 0.00460 |
| 2 | 0 | 0 | 0.00015 |
| 2 | 1 | 0 | 0.00010 |
| 2 | 1 | 1 | 0.00004 |
| 2 | 2 | 0 | -0.00007 |
| 2 | 2 | 1 | -0.00004 |
| 2 | 2 | 2 | -0.00002 |
| 3 | 0 | 0 | 0.00001 |

Table 4: *The largest values of $\omega(\vec{z})$. We have chosen $\alpha = \frac{1}{6}$ such that the classically perfect Polyakov loop is ultralocal in $d = 2$. Note the symmetry under permutations of the $z_i$.*

with the Fourier transform of $\delta^g(\vec{p}, 0) \Pi_d(\vec{p}, 0)^2 \Delta_d^g(\vec{p}, 0)^{-1}$. Again we restrict this function to the lattice points which in momentum space leads to

$$\sum_{\vec{l} \in \mathbf{Z}^{d-1}} \delta^g(\vec{p} + 2\pi\vec{l}, 0) \Pi_d(\vec{p} + 2\pi\vec{l}, 0)^2 \Delta_d^g(\vec{p} + 2\pi\vec{l}, 0)^{-1} =$$
$$[\Delta_d^g(\vec{p}, 0) - \alpha] \Delta_d^g(\vec{p}, 0)^{-1} = 1 - \alpha \Delta_d^g(\vec{p}, 0)^{-1}. \tag{4.15}$$

Because $\rho_{\mu\nu}(x - y)$ decays exponentially (as we saw in section 3) the resulting cut-off effects are again exponentially small. This is in agreement with a result obtained in [10] for a different renormalization group transformation. From eq.(4.14) one obtains an expression for the static quark-antiquark potential

$$V(\vec{r}) = -\frac{1}{(2\pi)^{d-1}} \int d^{d-1}p \; \delta^g(\vec{p}, 0)^2 \Pi_d(\vec{p}, 0)^2 \Delta_d^g(\vec{p}, 0)^{-1} \exp(i\vec{p} \cdot \vec{r}). \tag{4.16}$$

Note that $V(-\vec{r}) = V(\vec{r})$. First we consider the potential in $d = 2$. There the integral can be performed analytically and the potential takes the form

$$V(r) = \begin{cases} r/2 & \text{for} \quad r \geq 2 \\ -r^5/120 + r^4/12 - r^3/3 + 2r^2/3 - r/6 + 4/15 & \text{for} \quad 1 \leq r \leq 2 \\ r^5/40 - r^4/12 + r^2/3 + 7/30 & \text{for} \quad 0 \leq r \leq 1, \end{cases} \tag{4.17}$$

which is four times continuously differentiable. We have fixed an arbitrary constant in the potential such that it agrees with the continuum result $r/2$ at large distances. It is remarkable that the potential obtained from the classically perfect Polyakov loop then agrees exactly with the continuum potential for $r \geq 2$. The potential in $d = 2$ is shown in fig.5. Even for $1 \leq r \leq 2$ — and hence at any distance larger than one lattice spacing — it differs from the continuum result by less than 2 percent. It should be noted that the potential of the standard Polyakov loop agrees with the continuum result. However, it is restricted to lattice points, whereas the potential of the classically perfect Polyakov loop can be evaluated at any distance. The potential in $d = 4$ is compared to the continuum potential $-1/4\pi r$ in fig.6. We show the potential derived from the classically perfect Polyakov loop both along the space diagonal and along an axis. The two agree at the 1 percent level at any distance $r$, i.e. rotation invariance is restored with good accuracy. This is in contrast to the standard Wilson potential. Even at $r = 3$ its values at $\vec{r} = (3,0,0)$ and $(2,2,1)$ differ by more than 5 percent. The potential of the classically perfect Polyakov loop along the diagonal, i.e. $V(r/\sqrt{3}, r/\sqrt{3}, r/\sqrt{3})$, deviates from the continuum potential by less than 0.4 percent at any distance larger than 1.5 lattice spacings. In [10] the potential of the classically perfect Polyakov loop has been determined for a different renormalization group transformation at distances that correspond to lattice vectors. Even at $\vec{r} = (1,0,0)$ it agrees with the continuum answer at the percent level. This is not the case for our renormalization group transformation when we put $\alpha = 1/6$ providing ultralocality in $d = 2$. However, as an alternative criterion one may choose $\alpha$ such that $V(\vec{r})$ is as close as possible to the continuum potential. This would suggest $\alpha \approx 1/13$. For that value the agreement with the continuum potential is of the same quality as in [10]. In [14] we will systematically vary $\alpha$ and $\gamma$ to optimize both the potential and the locality of the action in $d = 4$.

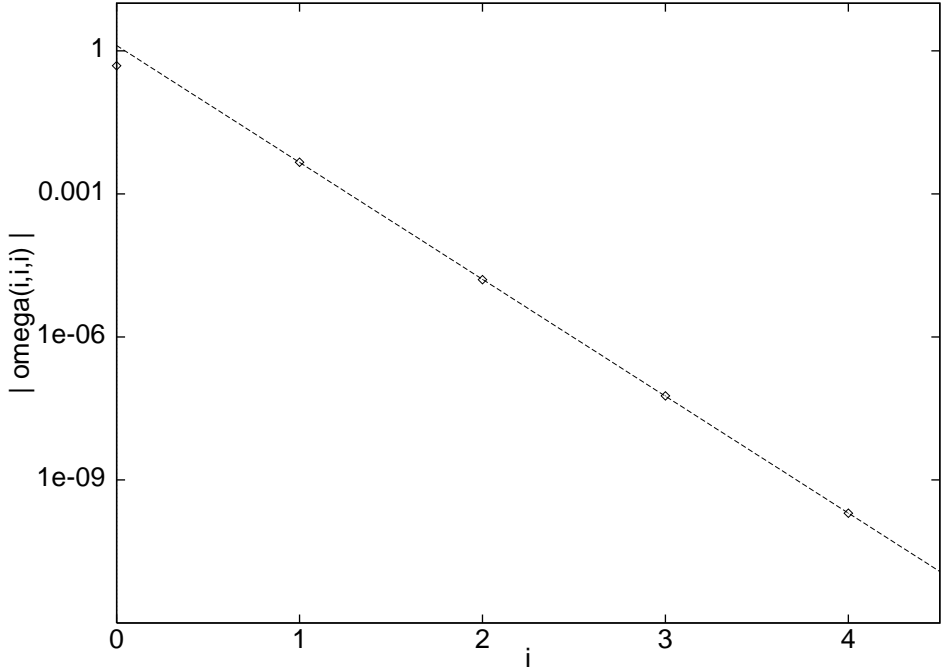

Figure 4: *Exponential decay of $|\omega(i,i,i)|$ for the classically perfect Polyakov loop (along the diagonal). The straight line is an exponential fit $|\omega(i,i,i)| \propto \exp(-ci)$ with $c = 5.646$ excluding the point $(0,0,0)$.*

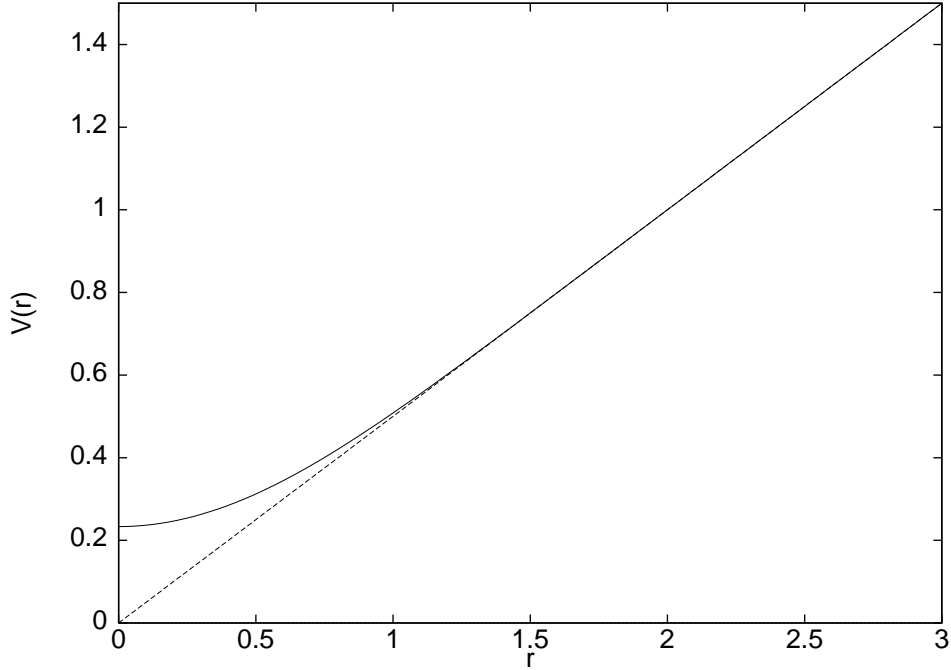

Figure 5: *The static quark-antiquark potential in $d = 2$ obtained from the classically perfect Polyakov loop (solid curve) compared to the continuum potential (dashed line).*

Next we turn to the currents. We start in the continuum and consider the current

$$j_\mu(x) = \bar\psi(x)\Gamma_\mu\psi(x), \tag{4.18}$$

where $\Gamma_\mu = \gamma_\mu$ for the vector current, and $\Gamma_\mu = \gamma_\mu\gamma_5$ for the axial vector current. Of course, more general expressions for $\Gamma_\mu$ are also possible; for example $\Gamma_\mu$ can be a flavor-nonsinglet current as well. In momentum space eq.(4.18) takes the form

$$j_\mu(p) = \frac{1}{(2\pi)^d} \int d^d q\ \bar\psi(p-q)\Gamma_\mu\psi(q). \tag{4.19}$$

We define lattice currents by inserting the classically perfect fields of eq.(4.1) into eq.(4.19) and we obtain

$$\begin{aligned}j_{\mu c}(p) &= \frac{1}{(2\pi)^d} \int_B d^d q \bar\Psi(p-q)\Delta^f(p-q)^{-1} \sum_{m\in\mathbb{Z}^d} \delta^f(p-q-2\pi m)\Gamma_\mu \\ &\quad \times\ \delta^f(q+2\pi m)\Pi(p-q-2\pi m)\Pi(q+2\pi m)\Delta^f(q)\Psi(q).\end{aligned} \tag{4.20}$$

We also want to define currents that are genuine lattice objects. Still they should have an exact interpretation in terms of continuum physics. A natural definition is the total flux of the continuum current through a $(d-1)$-dimensional face $f_{\mu,x}$ that separates two hypercubes $c_{x-\hat{\mu}/2}$ and $c_{x+\hat{\mu}/2}$. The corresponding lattice current $J_{\mu,x}$ associated with the link connecting the lattice points $x - \hat{\mu}/2$ and $x + \hat{\mu}/2$ is given by

$$J_{\mu,x} = \int_{f_{\mu,x}} d^{d-1}y \; j_\mu(y). \tag{4.21}$$

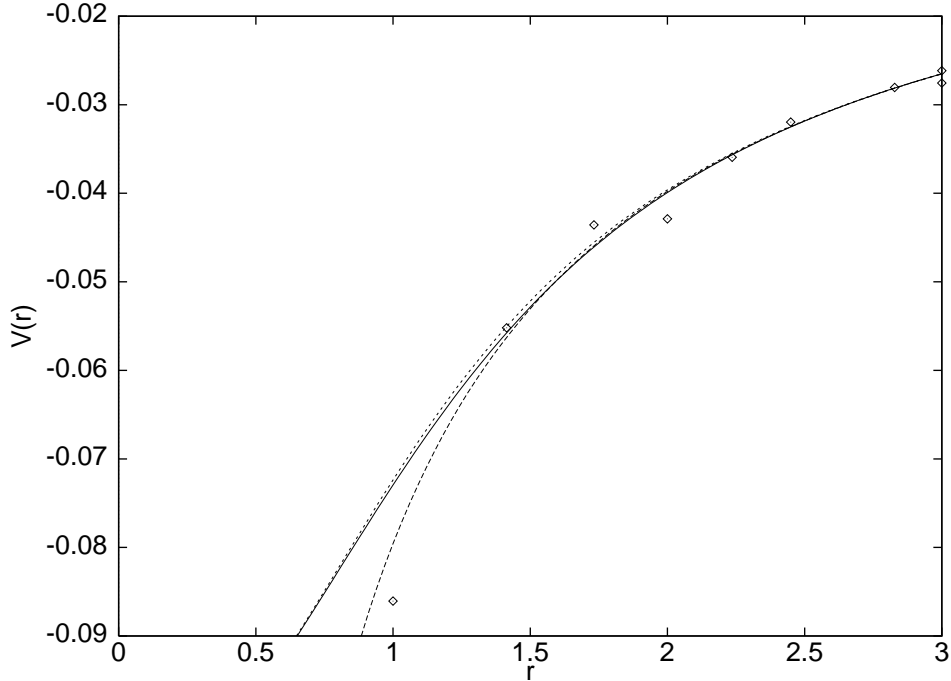

Figure 6: *The static quark-antiquark potential in $d = 4$. The dashed curve is the continuum Coulomb potential $-1/4\pi r$. The potential obtained from the classically perfect Polyakov loop along the space diagonal and along an axis is represented by the solid and dotted curve, respectively. The potential can be evaluated at any distance, whereas the potential of the standard Wilson theory, represented by the diamonds, is limited to lattice vectors.*

We call this the 'exact lattice current'. Its standard lattice divergence is by Gauss' law identical with the divergence of the continuum current integrated over a hypercube $c_x$

$$\delta J_x = \sum_\mu (J_{\mu,x+\hat{\mu}/2} - J_{\mu,x-\hat{\mu}/2}) = \int_{c_x} d^d y \; \partial_\mu j_\mu(y). \tag{4.22}$$

In momentum space the exact lattice current takes the form

$$J_\mu(p) = \sum_{l \in \mathbb{Z}^d} j_\mu(p + 2\pi l) \Pi_{\neg\mu}(p + 2\pi l)(-1)^{l_\mu}, \quad \Pi_{\neg\mu}(p) = \frac{p_\mu}{\hat{p}_\mu} \Pi(p). \tag{4.23}$$

The current is antiperiodic over the Brillouin zone in the $\mu$-direction because it lives on the link centers, which have a half-integer component $x_\mu$. As it stands the lattice current $J_\mu$ is expressed in terms of continuum fields. We now express it in terms of the lattice quark fields by using eq.(4.20)

$$\begin{aligned}
J_{\mu c}(p) &= \frac{1}{(2\pi)^d} \int_B d^d q\, \bar{\Psi}(p-q) \Delta^f(p-q)^{-1} \\
&\quad \times \sum_{l,m\in \mathbf{Z}^d} \delta^f(p+2\pi l - q - 2\pi m)\Gamma_\mu \delta^f(q+2\pi m)\Delta^f(q)^{-1}\Psi(q) \\
&\quad \times \Pi(p+2\pi l - q - 2\pi m)\Pi(q+2\pi m)\Pi_{\neg\mu}(p+2\pi l)(-1)^{l_\mu}. \quad (4.24)
\end{aligned}$$

The exact lattice current is analogous to the lattice field $\Psi$ because it is a genuine lattice object that still has an exact interpretation in terms of continuum physics. In particular its lattice divergence is the divergence of the continuum current averaged over a hypercube. This property was essential in a study of the axial anomaly in the Schwinger model [22]. It should be emphasized that correlation functions of the exact currents are not free of cut-off effects. In fact, their cut-off effects are not even exponentially small. Still the exact current is a natural object with useful properties in the perfect action approach to lattice field theory.

# 5 The perfect quark-gluon and 3-gluon vertex

In this section we switch on the interactions between quarks and gluons. However, we limit ourselves to the leading order in the gauge coupling $e$. Now the non-Abelian nature of the fields becomes important and we no longer suppress the color indices. The continuum quark field is denoted by $\bar{\psi}^i$, $\psi^i$ with $i \in \{1,2,...,N\}$ and the continuum $SU(N)$ gauge field is denoted by $a_\mu^a$ with $a \in \{1,2,...,N^2-1\}$. The gluon field strength is given by

$$f_{\mu\nu}^a(x) = \partial_\mu a_\nu^a(x) - \partial_\nu a_\mu^a(x) + ie f^{abc} a_\mu^b(x) a_\nu^c(x). \quad (5.1)$$

Here $f^{abc}$ are the structure constants of the $SU(N)$ algebra spanned by the Hermitean generators $\lambda^a$, such that $[\lambda^a, \lambda^b] = 2if^{abc}\lambda^c$ and $\text{Tr}(\lambda^a \lambda^b) = 2\delta_{ab}$. The continuum QCD action is given by

$$s[\bar{\psi},\psi,a] = \int d^d x \left\{ \frac{1}{4} f_{\mu\nu}^a(x) f_{\mu\nu}^a(x) + \bar{\psi}^i(x)[\gamma_\mu(\partial_\mu \delta_{ij} + ie a_\mu^a(x)\lambda_{ij}^a) + m]\psi^j(x) \right\}. \quad (5.2)$$

In momentum space the resulting quark-gluon vertex takes the form

$$v[\bar{\psi},\psi,a] = \frac{e}{(2\pi)^{2d}} \int d^d p\, d^d q\, \bar{\psi}^i(-p) i\gamma_\mu a_\mu^a(p-q)\lambda_{ij}^a \psi^j(q), \quad (5.3)$$

and the 3-gluon vertex is given by

$$v[a] = \frac{e}{(2\pi)^{2d}} \int d^d p\, d^d q\, f^{abc} a_\mu^a(-p)[p_\mu - q_\mu] a_\nu^b(p-q) a_\nu^c(q). \quad (5.4)$$

Again we want to integrate out the continuum fields and derive the perfect quark-gluon and 3-gluon vertex of the lattice theory. Before we can do so we need to define the renormalization group transformation up to $O(e)$. This can be done in many ways, and we do not want to make a specific choice yet. In general, for the quarks one can write

$$\Psi^i(p) = \sum_{l \in \mathbf{Z}^d} \psi^i(p + 2\pi l)\Pi(p + 2\pi l)$$
$$+ e \sum_{l \in \mathbf{Z}^d} \frac{1}{(2\pi)^d} \int d^dq \ K_\mu(p + 2\pi l, q + 2\pi l)a_\mu^a(p-q)\lambda_{ij}^a \psi^j(q + 2\pi l). \quad (5.5)$$

Here $K_\mu(p,q)$ is a regular kernel that specifies the renormalization group transformation. It describes how the continuum quark field in a hypercube $c_x$ is transported to the lattice point $x$. For a certain class of renormalization group transformations the kernel obeys an additional gauge covariance constraint. In these cases one requires that a continuum gauge transformation performed on $\psi$ and $a_\mu$ induces a lattice gauge transformation of $\Psi$. This implies the constraint $(p_\mu - q_\mu)K_\mu(p,q) = \Pi(p-q)\Pi(q) - \Pi(p)$. It is not really necessary to impose this constraint. In fact, in the next section we will select a specific renormalization group transformation for QCD for which the kernel is identically zero. To determine the kernel that corresponds to a given renormalization group transformation with a finite blocking factor $n$ one proceeds as follows. First one blocks from the continuum to a lattice of spacing 1 and performs one additional step of the block factor $n$ transformation. Then one derives a recursion relation for the kernel by demanding that the result agrees with the one obtained by blocking from the continuum to a lattice of spacing $n$.

Compared to the free fermion calculation of section 2 we now have an extra term

$$T[\bar\psi,\psi,a] = \frac{e}{(2\pi)^{2d}} \int d^dp d^dq \ [\bar\eta^i(-p)K_\mu(p,q)a_\mu^a(p-q)\lambda_{ij}^a \psi^j(q)$$
$$- \bar\psi^i(-p)K_\mu(-q,-p)a_\mu^a(p-q)\lambda_{ij}^a \eta^j(q)] \quad (5.6)$$

in the renormalization group transformation and hence in the exponential of eq.(2.5). The additional terms $v[\bar\psi,\psi,a]$ and $T[\bar\psi,\psi,a]$ modify the exponent $E[\bar\psi,\psi] + E[a]$ of the free theory to $E[\bar\psi,\psi,a] = E[\bar\psi,\psi] + E[a] + v[\bar\psi,\psi,a] + T[\bar\psi,\psi,a]$. Since the additional terms are $O(e)$ they change the equations of motion (and hence their solutions) to $\bar\psi_c + \delta\bar\psi_c$ and $\psi_c + \delta\psi_c$ and $a_c + \delta a_c$, where $\delta\bar\psi_c$, $\delta\psi_c$ and $\delta a_c$ are of $O(e)$. Inserting the new solution into the new exponent yields

$$E[\bar\psi_c + \delta\bar\psi_c, \psi_c + \delta\psi_c, a_c + \delta a_c] =$$
$$E[\bar\psi_c + \delta\bar\psi_c, \psi_c + \delta\psi_c] + E[a_c + \delta a_c] + v[\bar\psi_c,\psi_c,a_c] + T[\bar\psi_c,\psi_c,a_c] + O(e^2) =$$
$$E[\bar\psi_c,\psi_c] + E[a_c] + v[\bar\psi_c,\psi_c,a_c] + T[\bar\psi_c,\psi_c,a_c] + O(e^2) =$$
$$S[\bar\Psi,\Psi] + S[A] + V[\bar\Psi,\Psi,A] + O(e^2). \quad (5.7)$$

The essential observation is that $\bar\psi_c$, $\psi_c$ and $a_c$ are solutions of classical equations of motion, i.e. extrema of the corresponding exponents. This implies $E[\bar\psi_c + \delta\bar\psi_c, \psi_c +$ $\delta\psi_c] = E[\bar{\psi}_c, \psi_c] + O(e^2)$ and $E[a_c + \delta a_c] = E[a_c] + O(e^2)$. Hence it is sufficient to insert the old solutions of the equations of motion into the additional terms. This allows us to read off the perfect quark-gluon vertex

$$\begin{aligned} V[\bar{\Psi}, \Psi, A] &= v[\bar{\psi}_c, \psi_c, a_c] + T[\bar{\psi}_c, \psi_c, a_c] \\ &= \frac{1}{(2\pi)^{2d}} \int_{B^2} d^d p\, d^d q\ \bar{\Psi}^i(-p) e V_\mu(p,q) A_\mu^a(p-q) \lambda_{ij}^a \Psi^j(q). \end{aligned} \quad (5.8)$$

Using eqs.(2.9), (2.11) and (3.14) one identifies the quark-gluon vertex function as

$$\begin{aligned} V_\mu(p,q) &= \Delta^f(p)^{-1} \Delta_\mu^g(p-q)^{-1} \sum_{l,m \in \mathbf{Z}^d} \delta^g(p + 2\pi l - q - 2\pi m) \\ &\quad \times\ \Pi_\mu(p + 2\pi l - q - 2\pi m)(-1)^{l_\mu + m_\mu} \\ &\quad \times\ \Big[\delta^f(p + 2\pi l) i\gamma_\mu \delta^f(q + 2\pi m) \Pi(p + 2\pi l)\Pi(q + 2\pi m) \\ &\quad +\ K_\mu(p + 2\pi l, q + 2\pi m)\delta^f(q + 2\pi m)\Pi(q + 2\pi m) \\ &\quad -\ K_\mu(-q - 2\pi m, -p - 2\pi l)\delta^f(p + 2\pi l)\Pi(p + 2\pi l)\Big] \Delta^f(q)^{-1}. \quad (5.9) \end{aligned}$$

This expression is in the fixed point lattice Landau gauge. To transform it to a general gauge one proceeds in complete analogy to the free gluon case. We will do this explicitly in the next section when we choose a specific renormalization group transformation.

Also the renormalization group transformation for the gluons needs to be defined up to $O(e)$. We write

$$\begin{aligned} A_\mu^a(p) &= \sum_{l \in \mathbf{Z}^d} a_\mu^a(p + 2\pi l) \Pi_\mu(p + 2\pi l)(-1)^{l_\mu} \\ &\quad +\ e \sum_{l \in \mathbf{Z}^d} \frac{1}{(2\pi)^d} \int d^d q\ f^{abc} K_{\mu\nu\rho}(p + 2\pi l, q + 2\pi l) \\ &\quad \times\ a_\nu^b(p - q) a_\rho^c(q + 2\pi l)(-1)^{l_\mu}. \end{aligned} \quad (5.10)$$

Compared to the free gluon case of section 3 we now have an extra term

$$T[a] = \frac{ie}{(2\pi)^{2d}} \int d^d p\, d^d q\ f^{abc} D_\mu^a(-p) K_{\mu\nu\rho}(p,q) a_\nu^b(p-q) a_\rho^c(q) \quad (5.11)$$

in the exponential of the renormalization group transformation eq.(3.8). In complete analogy to the quark-gluon vertex one obtains for the 3-gluon vertex

$$\begin{aligned} V[A] &= v[a_c] + T[a_c] \\ &= \frac{1}{(2\pi)^{2d}} \int_{B^2} d^d p\, d^d q\ f^{abc} A_\mu^a(-p) e V_{\mu\nu\rho}(p,q) A_\nu^b(p-q) A_\rho^c(q). \quad (5.12) \end{aligned}$$

Using eq.(3.14) one identifies

$$\begin{aligned}
V_{\mu\nu\rho}(p,q) &= \Delta_\mu^g(p)^{-1}\Delta_\nu^g(p-q)^{-1}\Delta_\rho^g(q)^{-1} \sum_{l,m\in\mathbf{Z}^d} \delta^g(p+2\pi l - q - 2\pi m) \\
&\times \Pi_\nu(p+2\pi l - q - 2\pi m)(-1)^{l_\nu + m_\nu} \\
&\times \Big[\delta^g(p+2\pi l)[p_\mu + 2\pi l_\mu - q_\mu - 2\pi m_\mu]\delta_{\nu\rho}\delta^g(q+2\pi m) \\
&\times \Pi_\mu(p+2\pi l)\Pi_\rho(q+2\pi m)(-1)^{l_\mu + m_\rho} \\
&+ K_{\mu\nu\rho}(p+2\pi l, q+2\pi m)\delta^g(q+2\pi m)\Pi_\rho(q+2\pi m)(-1)^{m_\rho}\Big].
\end{aligned} \quad (5.13)$$

Also this expression is in the fixed point lattice Landau gauge.

# 6 A nonperturbative version of the renormalization group transformation

Before one uses perfect actions in numerical simulations of light quarks one would also like to incorporate nonperturbative effects. This requires numerical work along the lines of [4, 10]. Then it is essential to use a renormalization group transformation with a finite blocking factor $n$. In practice one would use $n = 2$. Below we work with general $n$. Our goal is to formulate a nonperturbative version of the renormalization group transformation whose perfect action reduces to the one of blocking from the continuum at least for weak fields. Then we know that the perfect action is very local (in lower dimensions even ultralocal), and therefore well suited for numerical simulations.

The nonperturbative formulation of lattice gauge theories works with compact link variables $U_{\mu,x} \in SU(N)$ connecting neighboring lattice points $x$ and $x+\hat{\mu}$. Note that here we have not associated $x$ with the link center; $x_\mu$ is now an integer. The parallel transporters are related to the gauge potential by

$$U_{\mu,x} = \exp(iA^a_{\mu, x+\hat{\mu}/2}\lambda^a). \quad (6.1)$$

As usual we have rescaled $eA_\mu \to A_\mu$. Under a gauge transformation $g$ the link variables transform as

$$^g U_{\mu,x} = g_x U_{\mu,x} g^+_{x+\hat{\mu}}. \quad (6.2)$$

In the renormalization group transformation we want to couple a blocked link $U'_{\mu,x'}$ on the coarse lattice to an average of $n^d$ parallel transporters on the fine lattice. The blocked link connects the centers of two neighboring $n^d$ blocks on the fine lattice. For each point $x \in x'$ in the first block we construct a parallel transporter as a product of $n$ consecutive links starting at $x$ and ending at the corresponding lattice point $x + n\hat{\mu}$ in the second block. Then we couple the parallel transporters to the blocked link $U'_{\mu,x'}$ by constructing a traceless anti-Hermitean matrix

$$B_{\mu,x'} = \frac{1}{n^d} \sum_{x \in x'} \log[U'_{\mu,x'}(\prod_{i=0}^{n-1} U_{\mu,x+i\hat{\mu}})^+] \qquad (6.3)$$

for each blocked link. A gauge transformation $g'$ on the coarse lattice

$$^{g'}U'_{\mu,x'} = g'_{x'} U'_{\mu,x'} g'_{x'+n\hat{\mu}}{}^+ \qquad (6.4)$$

induces a gauge transformation $g_x = g'_{x'}$ for $x \in x'$ on the fine lattice. Hence the anti-Hermitean matrix transforms as

$$^{g'}B_{\mu,x'} = g'_{x'} B_{\mu,x'} g'_{x'}{}^+. \qquad (6.5)$$

Note that $B_\mu$ transforms with the same $g'$ on both sides. Hence it is a charged vector field in the adjoint representation, not a lattice gauge field. In the renormalization group transformation we will need the covariant second derivative of $B_\mu$ in the $\mu$-direction

$$\Delta_\mu[U']B_{\mu,x'} = 2B_{\mu,x'} - U'_{\mu,x'} B_{\mu,x'+n\hat{\mu}} U'_{\mu,x'}{}^+ - U'_{\mu,x'-n\hat{\mu}}{}^+ B_{\mu,x'-n\hat{\mu}} U'_{\mu,x'-n\hat{\mu}}. \qquad (6.6)$$

(There is no summation over $\mu$.) Note that the fields are shifted by $n\hat{\mu}$ because the blocked lattice has lattice spacing $n$. Under a gauge transformation $g'$ the covariant second derivative transforms as

$$\Delta_\mu[^{g'}U']^{g'}B_{\mu,x'} = g'_{x'} \Delta_\mu[U'] B_{\mu,x'} g'_{x'}{}^+. \qquad (6.7)$$

Now we are ready to formulate the nonperturbative version of the renormalization group transformation for QCD. The effective action for the coarse variables after a factor $n$ renormalization group step is given by

$$\begin{aligned}\exp(-S'[\bar{\Psi}', \Psi', U']) &= \int \mathcal{D}\bar{\Psi}\mathcal{D}\Psi\mathcal{D}U \exp(-S[\bar{\Psi}, \Psi, U]) f[U] \\ &\times \exp\left\{-\sum_{x'} \frac{1}{a_n}(\bar{\Psi}'_{x'} - \frac{b_n}{n^d}\sum_{x \in x'}\bar{\Psi}_x)(\Psi'_{x'} - \frac{b_n}{n^d}\sum_{x \in x'}\Psi_x)\right\} \\ &\times \exp\left\{-\frac{1}{2e^2}\sum_{x',\mu}\text{Tr}\left[B^+_{\mu,x'}(\alpha_n + \gamma_n\Delta_\mu[U'])^{-1} B_{\mu,x'}\right]\right\}. (6.8)\end{aligned}$$

The functional $f[U]$ ensures the proper normalization of the partition function and is given by

$$f[U]^{-1} = \int \mathcal{D}U' \exp\left\{-\frac{1}{2e^2}\sum_{x',\mu}\text{Tr}\left[B^+_{\mu,x'}(\alpha_n + \gamma_n\Delta_\mu[U'])^{-1} B_{\mu,x'}\right]\right\}. \qquad (6.9)$$

For weak gauge fields the above transformation reduces to the discrete transformations of sections 2 and 3. Hence we know that the parameters

$$a_n = \frac{e^m - 1 - m}{m^2}\left(1 - \frac{1}{n}\right), \quad b_n = n^{(d-1)/2},$$
$$\alpha_n = \frac{1}{6}\left(1 - \frac{1}{n^2}\right), \quad \beta_n = n^{(d-2)/2}, \quad \gamma_n = -\frac{1}{72}\left(1 - \frac{1}{n^2}\right)^2 \qquad (6.10)$$

give a very local perfect action. From eq.(6.8) one automatically obtains $\beta_n = n^2$. ($\beta_n$ was introduced in eq.(3.27)). A fixed point is reached only for $\beta_n = n^{(d-2)/2}$ which limits us to $d = 4$. This is no surprise because only in $d = 4$ non-Abelian gauge fields are asymptotically free. Then the fixed point of the interacting theory is close to the Gaussian fixed point of the free theory.

It may seem that the above transformation is not gauge covariant, e.g. because the fine lattice quark field $\Psi_x$ is not properly parallel transported to the block center $x'$. This is, however, not required for the gauge invariance of the blocked action. In fact, performing a gauge transformation $g'$ on the coarse lattice and the induced gauge transformation $g$ with $g_x = g'_{x'}$ for $x \in x'$ on the fine lattice leads to

$$(^{g'}\bar{\Psi}'_{x'} - \frac{b_n}{n^d}\sum_{x\in x'}{}^g\bar{\Psi}_x)(^{g'}\Psi'_{x'} - \frac{b_n}{n^d}\sum_{x\in x'}{}^g\Psi_x) =$$
$$(\bar{\Psi}'_{x'}g'^{+}_{x'} - \frac{b_n}{n^d}\sum_{x\in x'}\bar{\Psi}_x g^{+}_x)(g'_{x'}\Psi'_{x'} - \frac{b_n}{n^d}\sum_{x\in x'}g_x\Psi_x) =$$
$$(\bar{\Psi}'_{x'} - \frac{b_n}{n^d}\sum_{x\in x'}\bar{\Psi}_x)(\Psi'_{x'} - \frac{b_n}{n^d}\sum_{x\in x'}\Psi_x). \qquad (6.11)$$

Using gauge invariance of the fine action and of the measure one observes that the coarse action is indeed gauge invariant. In particular, for the above transformation the kernel $K_\mu(p,q)$ of eq.(5.5) vanishes identically and the vertex function of eq.(5.9) reduces to

$$\begin{aligned}V_\mu(p,q) &= \Delta^f(p)^{-1}\Delta^g_\mu(p-q)^{-1}\sum_{l,m\in\mathbb{Z}^d}\delta^g(p+2\pi l - q - 2\pi m)\\ &\times \Pi_\mu(p+2\pi l - q - 2\pi m)(-1)^{l_\mu+m_\mu}\\ &\times \delta^f(p+2\pi l)i\gamma_\mu\delta^f(q+2\pi m)\Pi(p+2\pi l)\Pi(q+2\pi m)\Delta^f(q)^{-1}.(6.12)\end{aligned}$$

Here we are in the fixed point lattice Landau gauge. To go back to a general gauge we perform the gauge transformation of eq.(3.17) both on the fermions and on the gauge fields and obtain

$$\begin{aligned}V_\mu(p,q) &= \Delta^f(p)^{-1}\Delta^g_{\mu\nu}(p-q)^{-1}\sum_{l,m\in\mathbb{Z}^d}\delta^g(p+2\pi l - q - 2\pi m)\\ &\times \Pi_\nu(p+2\pi l - q - 2\pi m)(-1)^{l_\nu+m_\nu}\end{aligned}$$

$$\times \quad \delta^f(p+2\pi l)i\gamma_\nu \delta^f(q+2\pi m)\Pi(p+2\pi l)\Pi(q+2\pi m)\Delta^f(q)^{-1}$$
$$+ \quad \frac{(\widehat{p-q})_\mu \Delta_\mu^g(p-q)^{-1}}{\sum_\rho (\widehat{p-q})_\rho^2 \Delta_\rho^g(p-q)^{-1}}[\Delta^f(q)^{-1} - \Delta^f(p)^{-1}]. \tag{6.13}$$

This expression contains products of up to five $\gamma$-matrices. They may be decomposed into 1, $\gamma_\mu$, $\sigma_{\mu\nu}$, $\gamma_5$ and $\gamma_\mu \gamma_5$. The $\sigma_{\mu\nu}$-term represents a perfect 'clover' term. This decomposition will be carried out in [14]. There we will also parametrize the vertex function in terms of parallel transporters connecting the quark fields. Here we consider the vertex function for 2-d Abelian gauge fields. In addition we assume that the fermion fields are constant in the 2-direction, i.e. $\bar{\Psi}_{(x_1,x_2)} = \bar{\Psi}_{(x_1,0)}$, $\Psi_{(x_1,x_2)} = \Psi_{(x_1,0)}$. Hence they are effectively 1-dimensional. Then the vertex function can be computed analytically in coordinate space, and to $O(e)$ the total action takes the form

$$S[\bar{\Psi}, \Psi, A] = \sum_x \frac{1}{2} F_x^2$$
$$+ \left(\frac{m}{\hat{m}}\right)^2 \sum_x \left\{ \frac{1}{2}\bar{\Psi}_x \sigma_1 (1 + ieA_{1,x+\hat{1}/2})\Psi_{x+\hat{1}} - \frac{1}{2}\bar{\Psi}_{x+\hat{1}} \sigma_1 (1 - ieA_{1,x+\hat{1}/2})\Psi_x \right.$$
$$+ e^m \bar{\Psi}_x \Psi_x - \frac{1}{2}\bar{\Psi}_x(1 + ieA_{1,x+\hat{1}/2})\Psi_{x+\hat{1}} - \frac{1}{2}\bar{\Psi}_{x+\hat{1}}(1 - ieA_{1,x+\hat{1}/2})\Psi_x \Big\}$$
$$+ \left(\frac{m}{\hat{m}}\right)^2 \sum_{x,y,z} \left\{ \bar{\Psi}_x \sigma_2 \chi(x_1 - y_1, y_1 - z_1) ieA_{2,y}\Psi_z + \bar{\Psi}_x \sigma_3 \theta(x_1 - y_1, y_1 - z_1) eF_y \Psi_z \right\}.$$
$$\tag{6.14}$$

Here
$$F_x = A_{1,x-\hat{2}/2} + A_{2,x+\hat{1}/2} - A_{1,x+\hat{2}/2} - A_{2,x-\hat{1}/2} \tag{6.15}$$
is the field strength living at the plaquette centers. Note that in two dimensions
$$\sigma_{\mu\nu} F_{\mu\nu} = \frac{i}{2}[\sigma_\mu, \sigma_\nu] F_{\mu\nu} = -\sigma_3 F. \tag{6.16}$$

Hence the last term in eq.(6.14) represents a perfect clover term. The nonvanishing values of $\chi(x_1, y_1)$ and $\theta(x_1, y_1)$ are given in tables 5 and 6, and their relative weights are illustrated in figs.7 and 8. The vertex functions are peaked at $x_1 = -y_1$, i.e. where fermion and antifermion are at the same position. For fixed separation $x_1 + y_1$ the maximum occurs when the gauge field is close to the fermions, i.e. when $x_1 - y_1$ is small. It is remarkable that the vertex function is again ultralocal for 1-d fermions coupled to 2-d gauge fields, i.e. the coupling vanishes identically beyond short distances. The full 4-d vertex function will of course not be ultralocal. However, based on the experience with the propagators we expect it to be still very local.

By construction, in the classical limit $e \to 0$ the full renormalization group transformation reduces to the quadratic problem of section 3 even for arbitrarily strong abelian gauge fields. Since we know that for 2-d configurations the fixed point action is ultralocal at the quadratic level, it follows that for any 2-d abelian gauge field configuration the single plaquette Manton action [15] is classically perfect. This is a remarkable result, because it is valid for strong fields. Our construction of the renormalization group transformation appears very natural. The perfect action of the transformation reduces to the one of blocking from the continuum for weak fields. Its parameters can be chosen such that the perfect action is very local. Besides these nice analytic properties the transformation is well suited for numerical work. For example, in the classical limit $e \to 0$ (and for the values of $\alpha_n$ and $\gamma_n$ given in eq.(3.31)) it turns out that $f[U] = 1$. This is very useful in the nonperturbative numerical determination of the classically perfect action beyond the weak field limit. This will be discussed in a subsequent publication [14].

| $x_1$ | $y_1$ | $\chi(x_1, y_1)$ |
|---|---|---|
| -2 | 1 | 1/80 |
| -2 | 2 | 1/120 |
| -1 | 0 | 17/240 |
| -1 | 1 | 1/6 |
| 0 | 0 | 19/60 |

Table 5: *The nonvanishing values of the effectively 1-dimensional vertex function $\chi(x_1, y_1)$. Note that $\chi(x_1, y_1) = \chi(-x_1, -y_1) = \chi(y_1, x_1)$.*

| $x_1$ | $y_1$ | $\theta(x_1, y_1)$ |
|---|---|---|
| -1.5 | 0.5 | 1/80 |
| -1.5 | 1.5 | 1/120 |
| -0.5 | -0.5 | 7/120 |
| -0.5 | 0.5 | 19/120 |

Table 6: *The nonvanishing values of the effectively 1-dimensional vertex function $\theta(x_1, y_1)$ entering the perfect clover term. Note that $\theta(x_1, y_1) = \theta(-x_1, -y_1) = \theta(y_1, x_1)$.*

# 7 Conclusions

In this paper we have performed the analytic calculations that are necessary to construct a perfect action as well as perfect classical fields and their composite operators for full QCD. The technique of blocking from the continuum allowed us to derive closed analytic expressions for the perfect action up to $O(e)$. This is not the case for other renormalization group transformations. In general, one can only derive recursion relations for the perfect propagators and vertex functions, but one cannot iterate to the fixed point analytically. We are convinced that it is very useful to have closed analytic expressions for the fixed point quantities, not only because they provide physical insight, but also because they simplify the numerical calculations that are necessary to turn the analytic results into a practical lattice action that can be simulated.

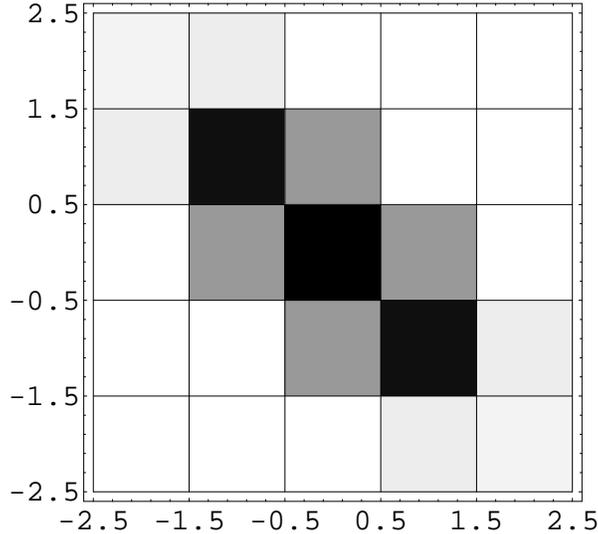

Figure 7: *Illustration of the relative weights of the nonvanishing values of the effectively 1-dimensional vertex function $\chi(x_1, y_1)$. Note the peak at $x_1 = -y_1$ where fermion and antifermion are at the same position.*

An alternative approach could start from the fixed point of free staggered fermions which has been constructed in [8, 25]. There one is limited to odd blocking factors, and one would use $n = 3$ [26]. Then also the blocking of the gauge field must be modified in a suitable way. Again one can work out the analytic expressions for the perturbatively perfect action. This is presently in progress [27].

For simulations it is essential that the renormalization group transformation we have used has a very local perfect action, provided its parameters are tuned appropriately. In particular, in lower dimensions (1-d for the fermions, 2-d for the gauge field) the perfect action of the free theory reduces to the ultralocal standard action. It is remarkable that even for arbitrarily strong 2-d abelian gauge fields the single plaquette Manton action is classically perfect. We expect that the fully nonperturbative perfect action of our renormalization group transformation will also be very local. Hence it should be possible to approximate it by a practically perfect action that can be simulated numerically.

The obvious next step is to extend our results nonperturbatively. This requires numerical work along the lines of [4, 10] which is presently in progress [14]. It is certainly nontrivial to include quarks in the nonperturbative perfect action. A natural first step will be to consider heavy quarks. First, their cut-off effects are large and therefore a perfect action should be much better than the standard Wilson action. Second, they are easier to simulate than light quarks, and third, the gauge interaction is weak at the scale of the heavy fermion mass such that our perturbatively perfect action may already be useful. In addition the perfect action of a heavy quark is more local than the one of a light quark. The ultimate goal is, of course, to include light fermions as well.

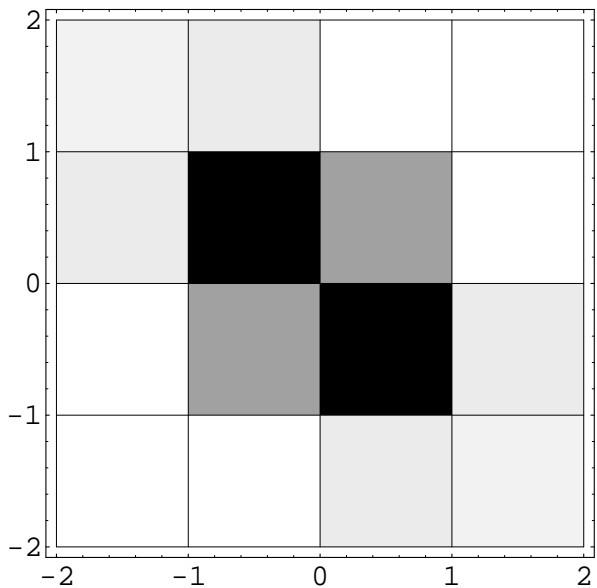

Figure 8: *Illustration of the relative weights of the nonvanishing values of the effectively 1-dimensional vertex function $\theta(x_1, y_1)$. Note that $x_1 + y_1$ is the separation between fermion and antifermion.*

## Acknowledgements

We are indebted to R. Brower who suggested that the perfect gauge action might be ultralocal in two dimensions. We thank him as well as T. DeGrand, P. Hasenfratz and F. Niedermayer for many stimulating discussions about various aspects of 'perfaction'. We have also benefited from discussions with T. Balaban, S. Chandrasekharan, A. Kronfeld, G. Palma and A. Tsapalis. One of the authors (UJW) likes to thank the institutes in Bern, Boulder and Erlangen for their hospitality during various stages of this work.